\global\def\draftcontrol{0}

   \def\versionno{ kw-charge}
\catcode`\@=11

\expandafter\ifx\csname draftcontrol\endcsname\relax\global\def\draftcontrol{0}
\fi

{\count255=\time\divide\count255 by 60
\xdef\hourmin{\number\count255}
\multiply\count255 by-60\advance\count255 by\time
\xdef\hourmin{\hourmin:\ifnum\count255<10 0\fi\the\count255}}
\def\draftdate{\number\month/\number\day/\number\year\ \ \ \hourmin }

\newcommand\makepapertitle{\par
  \begingroup
    \renewcommand\thefootnote{\@fnsymbol\c@footnote}%
    \def\@makefnmark{\rlap{\@textsuperscript{\normalfont\@thefnmark}}}%
    \long\def\@makefntext##1{\parindent 1em\noindent
            \hb@xt@1.8em{%
                \hss\@textsuperscript{\normalfont\@thefnmark}}##1}%
     \newpage
     \global\@topnum\z@   % Prevents figures from going at top of page.
     \@makepapertitle
     \thispagestyle{empty}\@thanks
  \endgroup
  \setcounter{footnote}{0}%
  \global\let\thanks\relax
  \global\let\makepapertitle\relax
  \global\let\@makepapertitle\relax
  \global\let\@thanks\@empty
  \global\let\@author\@empty
  \global\let\@date\@empty
  \global\let\@title\@empty
  \global\let\title\relax
  \global\let\author\relax
  \global\let\date\relax
  \global\let\and\relax
  \def\version{\let\version\@version\@gobble}
}
\def\@makepapertitle{%
  \newpage
   \ifnum\draftcontrol=1 {}
   \version\versionno
   \vskip 3em%
   \else
   \hfill\hbox to 3cm {\parbox{4cm}{\@pubnum}\hss}%
   \vskip 3em%
   \fi
   \begin{center}%
   \let \footnote \thanks
     {\LARGE {\@title}}%
     \vskip 1.5em%
     {\normalsize%\large
       \lineskip .5em%
       \begin{tabular}[t]{c}%
         \@author
       \end{tabular}\par}%
     \vskip 1.5em%
     {\@bstract}%
     \end{center}%
     \vskip 1.5em
     \@date%
   \par
}

\gdef\@pubnum{}
\def\pubnum#1{%
  \gdef\@pubnum{#1}}

\gdef\@bstract{}
\def\Abstract#1{%
  \gdef\@bstract{%
   \parbox{\textwidth-0pc}{%
   \centerline{\bf Abstract}\penalty1000%
\kern.2cm%
\noindent%\abstractfont \baselineskip=12pt
\renewcommand\baselinestretch{1.0}%
{#1}}}
}

\def\ps@paper{\let\@mkboth\@gobbletwo%
     \ifnum\draftcontrol=1
    \def\@oddfoot{\hbox to \textwidth{\tiny \versionno \hfil\tiny\draftdate}%
    \hskip -\textwidth \hbox to \textwidth{\hfil\rm\thepage\hfil}}%
     \else\def\@oddfoot{\hbox to \textwidth{\hfil\rm\thepage\hfil}}
     \fi
     \let\@evenfoot\@oddfoot
}
\def\body{\clearpage
          \pagestyle{paper}
    }
\def\@version#1{\ifnum\draftcontrol=1
\typeout{}\typeout{#1}\typeout{}
\vskip3mm\centerline{\hbox{\fbox{\normalsize{\tt DRAFT -- #1 -- }
                   {\draftdate}}}}\vskip3mm
\fi}
\let\version\@version
\long\def\eqlabel#1{\ifnum\draftcontrol=1
                    \tag@false  % there are some problems with multline without this
                    \tag*{(\theequation) \hbox to -0.2cm{\hspace{0cm}\small{#1}\hss}}
                    \refstepcounter{equation}
                    \edef\@currentlabel{\theequation}
                    \ltx@label{#1}          % use old LaTeX \label instead of new definition
                    \else
                    \label{#1}
                    \fi
                    }
\let\st@bibitem\@bibitem
\let\st@lbibitem\@lbibitem
\ifnum\draftcontrol=1
  \def\@bibitem#1{%
    \st@bibitem{#1}\a@@label{#1}\ignorespaces}
  \def\@lbibitem[#1]#2{%
    \st@lbibitem[#1]{#2}\a@@label{#2}\ignorespaces}
  \def\a@@label#1{%
    \gdef\a@lab{\smash{\normalfont\small#1}}
    \ifvmode
      \if@inlabel
        \global\setbox\@labels\hbox{%
          \llap{\a@lab\let\a@lab\relax
                \kern\@totalleftmargin\kern\marginparsep}%
          \box\@labels}%
      \fi
    \fi}
\fi
\documentclass[12pt,letterpaper]{article}
\usepackage{amsmath,amssymb,array,calc,epsfig,rotating,bm}
\usepackage[sort]{cite}
\usepackage{graphicx}
\usepackage{psfrag,verbatim}
\usepackage{xcolor}
\usepackage{hyperref}

\ifnum\draftcontrol=1
\fi
\renewcommand\baselinestretch{1.25}
\renewcommand\section{\@startsection {section}{1}{\z@}%
                                   {-3.5ex \@plus -1ex \@minus -.2ex}%
                                   {2.3ex \@plus.2ex}%
                                   {\normalfont\large\bfseries}}
\renewcommand\subsection{\@startsection{subsection}{2}{\z@}%
                                   {-3.25ex\@plus -1ex \@minus -.2ex}%
                                   {1.5ex \@plus .2ex}%
                                   {\normalfont\normalsize\bfseries}}
\renewcommand\subsubsection{\@startsection{subsubsection}{3}{\z@}%
                                   {-3.25ex\@plus -1ex \@minus -.2ex}%
                                   {1.5ex \@plus .2ex}%
                                   {\normalfont\normalsize\it}}
\renewcommand\paragraph{\@startsection{paragraph}{4}{\z@}%
                                   {-3.25ex\@plus -1ex \@minus -.2ex}%
                                   {1.5ex \@plus .2ex}%
                                   {\normalfont\normalsize\bf}}

\numberwithin{equation}{section}

\def\revise#1       {\raisebox{-0em}{\rule{3pt}{1em}}%
                     \marginpar{\raisebox{.5em}{\vrule width3pt\
                     \vrule width0pt height 0pt depth0.5em
                     \hbox to 0cm{\hspace{0cm}{%
                     \parbox[t]{4em}{\raggedright\footnotesize{#1}}}\hss}}}}

\newcommand\nxt[1]  {\\\fnxt#1}
\newcommand{\ie}{{\it i.e.,}\ }

\def\cala         {{\cal A}}

\def\cale         {{\cal E}}

\def\call         {{\cal L}}
\def\calm         {{\cal M}}
\def\caln         {{\cal N}}
\def\calo         {{\cal O}}
\def\calp         {{\cal P}}

\def\cals         {{\cal S}}

\def\calv         {{\cal V}}

\def\reals        {{\mathbb R}}
\def\zet          {{\mathbb Z}}

\def\del          {\partial}

\def\tr           {\mathop{\rm Tr}}
\def\Re           {{\rm Re\hskip0.1em}}
\def\Im           {{\rm Im\hskip0.1em}}

\def\sqr#1#2{{\vcenter{\vbox{\hrule height.#2pt
 \hbox{\vrule width.#2pt height#1pt \kern#1pt
 \vrule width.#2pt}\hrule height.#2pt}}}}

\newcommand{\ft}[2]{{\textstyle{\frac{#1}{#2}}}}

\def\aa1{\phi}
\def\cc1{\psi}

\def\arctanh{{\rm arctanh}}

\def\Om{\Omega}

\catcode`\@=12

\begin{document}

%%%
%%%%%% text starts here
%%%%%%%%%

\title{\bf Fluxification and scalarization of the conifold black holes}

\date{January 12, 2025}
%\date\today

\author{
Alex Buchel\\[0.4cm]
\it Department of Physics and Astronomy\\ 
\it University of Western Ontario\\
\it London, Ontario N6A 5B7, Canada\\
\it Perimeter Institute for Theoretical Physics\\
\it Waterloo, Ontario N2J 2W9, Canada\\
}

\Abstract{
We study thermal states of the strongly coupled Klebanov-Witten
${\cal N}=1$ superconformal gauge theory with R-symmetry chemical
potential. The theory has two distinct dimension $\Delta=3$ chiral
primary operators that develop a condensate at the same critical value
of the chemical potential. One instability is associated with the
spontaneous {\it fluxification} (the appearance of the 3-form flux) of
the holographic dual black hole, while the other one is associated
with the spontaneous {\it scalarization} (the activation of the
deformation mode of the conifold).  Different phases dominate in grand
canonical and microcanonical ensembles.
}

\makepapertitle

\body

\version\versionno
\tableofcontents

\section{Introduction and summary}\label{intro}

Klebanov-Witten model \cite{Klebanov:1998hh} (KW) is an important example of the holographic correspondence between  
strongly coupled $\caln=1$ superconformal $SU(N)\times SU(N)$  quiver gauge theory
and the near-horizon limit of D3 branes on the conical singularity of type IIB string theory. 
This theory arises as an IR limit of the renormalization group flow of the $\zet_2$ orbifold of
maximally supersymmetric $\caln=4$ supersymmetric Yang-Mills theory
\cite{Maldacena:1997re,Douglas:1996sw,Kachru:1998ys}, triggered by the mass term to the
chiral multiplet of $\caln=2$ vector hypermultiplet of the parent UV
theory\footnote{See \cite{Buchel:2021yay} for a recent discussion of the KW model in the holographic context.}. 

In this paper we study (via holography \cite{Witten:1998zw}) the thermal states of the KW gauge theory plasma, charged
under $U(1)$ R-symmetry. In 10d type IIB supergravity these states are described by  spinning black  holes
on the conifold \cite{Candelas:1989js}; and from the effective 5d perspective, after the Kaluza-Klein reduction
on the conifold base, they are represented by the Reissner-Nordstrom black holes in asymptotically
$AdS_5$ space-time. There is a trivial R-symmetry preserving phase of the model with the Gibbs free
energy density given by
\begin{equation}
\Omega =-\frac{c_{KW}}{2\pi^2}\left(\alpha^4+\frac 19\alpha^2 \mu^2\right)\,,\qquad
\frac{T}{\mu}=\frac{18\alpha^2-\mu^2}{18\pi\mu}\,,
\eqlabel{omsysi}
\end{equation}
where $T$ and $\mu$ are the temperature and the $U(1)$ chemical potential correspondingly;
$c_{KW}$ is the central charge of the KW theory, and $\alpha$ is an arbitrary auxiliary
scale\footnote{This scale can be eliminated in favor  of $\frac T\mu$, but then the thermodynamic
formulas look unnecessarily complicated.}. Note that the symmetric phase has an extremal limit,
\ie  as $\frac T\mu\to 0$, with the finite entropy density at the extremality.
But this can not be the full story: it has been known for a long time \cite{Ceresole:1999zs}
that the KW gauge theory has a rich spectrum of R-charged operators, which can potentially condense at low
temperatures (before the extremality is reached and quantum effects would become important \cite{Turiaci:2023wrh}).
Condensation of charged operators would spontaneously break the $U(1)$ R-symmetry, leading
to interesting superconducting phases of the model \cite{Hartnoll:2008vx}. In the dual gravitational
description, the holographic black holes, representing these superconducting states, develop 'hair'.
Construction of the corresponding holographic superconducting phases in the KW model thus far proved
elusive --- the complexity of the conifold hairy black holes presented here could be the reason.

We delegate all the technical details to section \ref{tech}, and summarize now the results and
point the possible future directions.

Our main players are the two chiral primary operators of the KW gauge theory:
$\calo_F\equiv \tr(W_1^2+W_2^2)$ and $\calo_S\equiv \tr(W_1^2-W_2^2)$, where $W_i$ are the gauge superfields
corresponding to the two gauge group factors of $SU(N)\times SU(N)$ quiver. While being very similar
from the gauge theory perspective, the operators have a very different encoding in the
holographic dual\footnote{This explains the subscripts
$F$ (flux) and $S$ (scalar) in these operators.}: $\calo_S$ is realized as a mode deforming the base of
the conifold, while $\calo_F$ is represented by a certain combination of 3-form NSNS and RR fluxes.  
Both operators have the same R-charge, $R(\calo_{S})=R(\calo_F)=2$, and thus the same conformal
dimension $\Delta=\frac 32 R$. Because of this, they condense at the same\footnote{We believe this is the first example in the literature
of the holographic superconductor with several distinct operators condensing simultaneously.} critical value of $\frac T\mu$,
\begin{equation}
\frac{T}{\mu}\bigg|_{crit}=0.030338(3)\,.
\eqlabel{tmucin}
\end{equation}
From the dual gravitational perspective, condensation of $\calo_S$ leads to scalarization of a spinning
conifold black hole, while the condensation of  $\calo_F$ leads to a black hole { fluxification}.

\begin{figure}[ht]
\begin{center}
\psfrag{z}[tt][][1.0][0]{{$T/\mu$}}
\psfrag{w}[tt][][1.0][0]{{$\hat\cale/(\hat\rho)^{4/3}$}}
\psfrag{x}[bb][][1.0][0]{{$\hat\Omega/\mu^4$}}
\psfrag{y}[tt][][1.0][0]{{$\hat\cals/\hat\rho$}}
\includegraphics[width=3in]{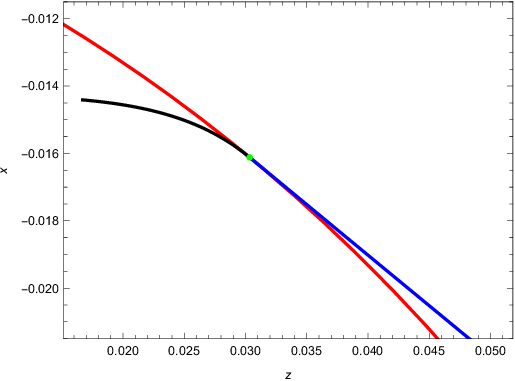}
\includegraphics[width=3in]{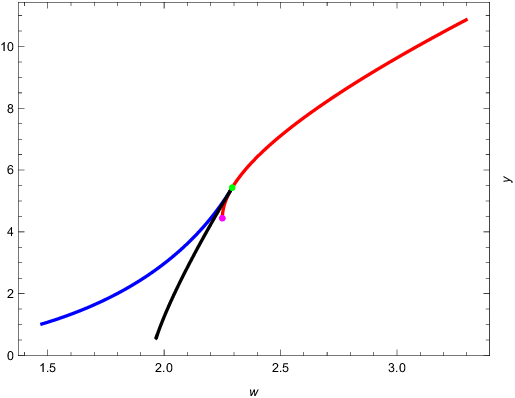}
\end{center}
  \caption{The phase diagram of the KW gauge theory in the grand canonical ensemble (the left panel)
  and the microcanonical ensemble (the right panel). The red curves represent $U(1)$-symmetric phase
  of the model; the phases with $\langle\calo_S\rangle\ne 0$ are in blue, and the phases with $\langle\calo_F\rangle\ne 0$ are in black.
The green dot indicates the critical point \eqref{tmucin}. The magenta dot is the extremal limit of $AdS_5$
Reissner-Nordstrom black hole.
} \label{figure2}
\end{figure}

The phase diagram of the KW model in the near-critical region is shown in fig.~\ref{figure2}. 
The left panel shows the phase diagram in the grand canonical ensemble,
and the right panel in the microcanonical ensemble.  The red curves represent the
R-symmetry preserving phase, see \eqref{omsysi}. We use $\ \hat{}\ $ notation in all the thermodynamic densities
(the Gibbs free energy $\hat\Omega$, the energy  $\hat\cale$, the entropy  $\hat\cals$ and the charge
$\hat\rho$ ) to
remove the central charge factor $\frac{c_{KW}}{2\pi^2}$, see \eqref{thermo}. The phases of the model with the
$\langle\calo_S\rangle\ne 0$ and  $\langle\calo_F\rangle=0$ condensates are in blue, and those with  $\langle\calo_S\rangle= 0$ and
$\langle\calo_F\rangle\ne 0$  are in black.
The green dot represents the critical point \eqref{tmucin}, and the magenta dot is the extremality of the symmetric phase. 
In the grand canonical ensemble, the flux-hair phase is a dominant one, while in the microcanonical
ensemble it is the scalar-hair phase that is a dominant one. The scalar-hair phase in the grand canonical
ensemble is an 'exotic' phase \cite{Buchel:2009ge,Buchel:2017map,Buchel:2018bzp}
--- it originates at the criticality, however it extends to higher, rather than the lower temperatures.

What are the possible extensions?
\begin{itemize}
\item We studied either the purely scalarized or the purely fluxified phases of the KW model; however it is possible
that there exist superconducting phases with both $\langle\calo_S\rangle\ne 0$ and  $\langle\calo_F\rangle\ne 0$.
Such phases would interpolate between the $S$ and $F$ phases discussed. 
These phases can be explored within the effective action described in this paper.
\item Related to above, phenomenology of holographic superconductors with multiple
simultaneous condensates is unexplored.
\item KW model is conformal. However, we can add fractional D3 branes and turn it into Klebanov-Strassler (KS) model
\cite{Klebanov:2000hb}. The construction of the corresponding gravitational dual goes unchanged, except
we need to keep $q\ne 0$, $p=0$ in \eqref{seff},  and of the two truncations in \eqref{backw2} only $\bm{(1)}$
is consistent. Analysis of this model is an opportunity to investigate the interplay between the spontaneous chiral symmetry breaking in the
KS model and the R-symmetry chemical potential.
\end{itemize}

\section{Technical details}\label{tech}

\subsection{Effective action and consistent truncations}

The starting point in constructing the relevant effective holographic action is the 
$SU(2)\times SU(2)$ invariant consistent truncation of type IIB supergravity
of warped deformed conifold \cite{Candelas:1989js} with fluxes, constructed
in\footnote{We use notations of \cite{Buchel:2014hja}.}
\cite{Cassani:2010na} (CF). We will need a consistent five dimensional $\caln=2$ gauged
supergravity subtruncation of the CF model (identified in \cite{Cassani:2010na})
that contains Klebanov-Witten \cite{Klebanov:1998hh} model: 
\begin{equation}
S_{eff}=\frac{1}{2\kappa_5^2} \int_{\calm_5} R \star 1+S_{kin,scal}+S_{kin,vect}+S_{top}+S_{pot}\,, 
\eqlabel{seff}
\end{equation}
with 
\begin{equation}
\begin{split}
&S_{kin,scal}=-\frac{1}{2\kappa_5^2}\int_{\calm_5}\biggl\{
\frac{28}{3} du^2+\frac 43 dv^2+\frac 83 dudv +d\tau^2+
\sinh^2\tau\ (d\theta-3 A)^2\\
&+e^{-4u-\phi}\biggl[\cosh(2\tau)\ (h_1^\Phi)^2+\cosh^2\tau|h_1^\Om|^2-\sinh^2\tau\ \Re(e^{-2i \theta}(h_1^\Om)^2)
\\& +2\sinh(2\tau)\ h_1^\Phi\ \Re(i e^{-i \theta}h_1^\Om)
\biggr]
+e^{-4u+\phi}\biggl[h\to g\biggr]+\frac 12 d\phi^2+\frac 12 e^{2\phi} dC_0^2+2 e^{-8 u}f_1^2
\biggr\} \star 1\,,
\end{split}
\eqlabel{sks}
\end{equation}
\begin{equation}
\begin{split}
&S_{kin,vect}=-\frac{1}{2\kappa_5^2}\int_{\calm_5}\biggl\{
\frac 12 e^{\ft 83 u +\ft 83 v}\ (dA)^2+
e^{-\ft 43 u -\ft 43 v }\ (da_1^J)^2
\biggr\} \star 1\,,
\end{split}
\eqlabel{svect}
\end{equation}
\begin{equation}
\begin{split}
&S_{top}=\frac{1}{2\kappa_5^2}\int_{\calm_5}\ A\wedge da_1^J\wedge da_1^J\,,
\end{split}
\eqlabel{stop}
\end{equation}
\begin{equation}
\begin{split}
&S_{pot}=\frac{1}{2\kappa_5^2}\int_{\calm_5}\biggl\{
e^{-\ft 83 u -\ft 23 v}\ R_{T^{1,1}}
-2e^{-\ft{32}{3}u-\ft 83 v} f_0^2-e^{-\ft{20}{3}u-\ft 83 v-\phi} \biggl[\Re(-e^{-2i\theta}\sinh^2\tau (h_0^\Om)^2
\\&+2i p  e^{-i \theta}\sinh(2\tau) h_0^\Om)
+\cosh^2\tau |h_0^\Om|^2+p^2 \cosh(2\tau)
\biggr]\\
&-e^{-\ft{20}{3}u-\ft 83 v+\phi}\biggl[
h\to g\,, p\to (q-p C_0)
\biggr]
\biggr\} \star 1\,,
\end{split}
\eqlabel{spot}
\end{equation}
\begin{equation}
R_{T^{1,1}}=-4 e^{-4 u +2 v}+24 e^{-2u}\cosh\tau-9 e^{-2 v}\ \sinh^2 \tau\,.
\eqlabel{rt11}
\end{equation}
Various fields in \eqref{seff} uplifts to 10d type IIB supergravity as follows:
\begin{itemize}
\item The 10d Einstein frame metric is a direct warped product of metric on $\calm_5$ and
a metric on the deformed coset
\begin{equation}
T^{1,1}\equiv \frac{SU(2)\times SU(2)}{U(1)}\,,
\eqlabel{deft11}
\end{equation}
parameterized by four 0-forms $\{u,v,\tau,\theta\}$ and a single 1-form $A$ on $\calm_5$,
\begin{equation}
\begin{split}
&ds_{10}^2=e^{-\frac 83 u-\frac 23 v}\cdot \underbrace{\quad ds_5^2\quad }_{{\rm metric\ on\ }\ \calm_5}\ +
\underbrace{\sum_{I=1}^5 E^I E^I}_{{\rm metric\ on\ }\ T^{1,1}}\,,\\
&\sum_{I=1}^5 E^I E^I\equiv \frac16 e^{2u}\cosh\tau\biggl(e_1^2+e_2^2+e_3^3+e_4^2\biggr)
\\&+\frac 13 e^{2u}\sinh\tau\biggl((e_1e_3+e_2e_4)\cos\theta+(e_1e_4-e_2e_3)\sin\theta\biggr)+\frac19e^{2v}(e_5-3A)^2\,,
\end{split}
\end{equation}
where $e_i$ are the standard coframe 1-forms on $T^{1,1}$ \cite{Minasian:1999tt},
\begin{equation}
\begin{split}
&e_1=-\sin\theta_1\ d(\phi_1)\,,\qquad e_2=d(\theta_1)\,,\\
&e_3=\cos\psi\ \sin\theta_2\ d(\phi_2)-\sin\psi\ d(\theta_2)\,,\\
&e_4=\sin\psi\ \sin\theta_2\ d(\phi_2)+\cos\psi\ d(\theta_2)\,,\\
&e_5=d(\psi)+\cos\theta_1\ d(\phi_1)+\cos\theta_2 d(\phi_2)\,, 
\end{split}
\eqlabel{coframe}
\end{equation} 
for angular coordinates $\{\theta_1,\phi_1,\theta_2,\phi_2,\psi\}$ with ranges
$0\le \theta_{1,2}<\pi$, $0\le \phi_{1,2}<2\pi$, and $0\le \psi <4\pi$.
Here we first encounter the complex scalar \colorbox{yellow}{$\tau e^{i\theta}$},
which condensation would lead to { scalarization} of the conifold black holes.
\item $\phi$ and $C_0$ are type IIB dilaton and axion.
\item To parameterize NSNS and RR 3-form fluxes we introduce left-invariant 1- and 2-forms
on the coset,
\begin{equation}
\begin{split}
&\eta=-\frac 13 e_5\,,\qquad \Omega=\frac 16 (e_1+i e_2)\wedge (e_3-i e_4)\,,\\
&J=\frac 16 (e_1\wedge e_2-e_3\wedge e^4)\,,\qquad \Phi=\frac{1}{6}(e_1\wedge e_2+e_3\wedge e_4)\,. 
\end{split}
\eqlabel{lif}
\end{equation}
\nxt NSNS 2-form potential $B_2$ and a 3-form flux $H_3$ are parameterized by a constant $p$,
a real 0-form $b^\Phi$, and a complex 0-form $b^\Omega$ on $\calm_5$:
\begin{equation}
\begin{split}
H_3=&p\ \Phi\wedge \eta+ d(B_2)\,,\qquad B_2=\Re(b^\Om\Om)+b^\Phi\Phi\,.
\end{split}
\eqlabel{defb2h3}
\end{equation}
The field strength $H_3$ can be decomposed in a basis of left-invariant forms on $T^{1,1}$ \eqref{lif}:
\begin{equation}
\begin{split}
H_3=&\Re\left[h_1^\Om\wedge \Om+h_0^\Om\ \Om\wedge (\eta+A)\right]
+h_1^\Phi\wedge \Phi+p\ \Phi\wedge (\eta+A)\,,
\end{split}
\eqlabel{h3decompose}
\end{equation}
where we defined
\begin{equation}
\begin{split}
&h_1^\Om=db^\Om-3i\ A\ b^\Om\equiv Db^\Om\,,\qquad h_0^\Om=3 i\ b^\Om\,,\qquad h_1^\Phi=db^\Phi-p\ A\equiv Db^\Phi\,.
\end{split}
\eqlabel{shdef}
\end{equation}
As we will see, the mode responsible for { fluxification} of the conifold black holes
is \colorbox{yellow}{$\Im(b^\Om)$}.
\nxt RR 2-form potential $C_2$ and a 3-form flux $F_3$ are parameterized by a constant $q$,
a real 0-form $c^\Phi$, and a complex 0-form $c^\Omega$ on $\calm_5$:
\begin{equation}
\begin{split}
F_3=q\ \Phi\wedge \eta+ d(C_2)-C_0 H_3\,,\qquad C_2=\Re(c^\Om\Om)+c^\Phi\Phi\,.
\end{split}
\eqlabel{defc2g3}
\end{equation}
The field strength $F_3$ can be decomposed in a basis of left-invariant forms on $T^{1,1}$ \eqref{lif}:
\begin{equation}
\begin{split}
F_3=&\Re\left[g_1^\Om\wedge \Om+g_0^\Om\ \Om\wedge (\eta+A)\right]\\
&+g_1^\Phi\wedge \Phi+(q-C_0 p)\ \Phi\wedge (\eta+A)\,,
\end{split}
\eqlabel{g3decompose}
\end{equation}
where we defined 
\begin{equation}
\begin{split}
&g_1^\Om=dc^\Om-3i\ A\ c^\Om-C_0 Db^\Om\equiv Dc^\Om-C_0 Db^\Om\,,\qquad g_0^\Om=3 i\ (c^\Om-C_0 b^\Om)\,,\\
&g_1^\Phi=dc^\Phi-q\ A-C_0 Db^\Phi\equiv Dc^\Phi-C_0Db^\Phi\,. 
\end{split}
\eqlabel{sgdef}
\end{equation}
The RR flux mode \colorbox{yellow}{$\Re(c^\Om)$} will take part in { fluxification} as well.
\item The remaining fields of the effective action \eqref{seff}, \ie the 0-form $f_0$ (in $S_{pot}$,
see \eqref{spot})
\begin{equation}
\begin{split}
f_0=k+p c^\Phi-q b^\Phi+3 \Im\left[b^\Om\overline{c^\Om}\right]\,,
\end{split}
\eqlabel{f0def}
\end{equation}
and the 1-form $f_1$ (in $S_{kin,scal}$,
see \eqref{sks})
\begin{equation}
\begin{split}
f_1&=d(a)-2a_1^J-kA+\frac 12 (q b^\Phi-p c^\Phi) A+\frac 12 \biggl[
-b^\Phi Dc^\Phi+\Re\left[b^\Om\overline{Dc^\Om}\right]-b\leftrightarrow c
\biggr]\,,
\end{split}
\eqlabel{f1def}
\end{equation}
are parameterized by a constant $k$, and additional  0-form $a$ and a 1-form $a_1^J$.
$f_0$ and $f_1$ describe the self-dual 5-form field strength of type IIB
supergravity\footnote{For detailed discussion see section 2.4 of \cite{Buchel:2014hja}.}.
\item The 5d gravitational coupling $\kappa_5$ is related to the UV central charge $c_{UV}$ of
the boundary gauge theory as follows:
\begin{equation}
\kappa_5^2=\frac{\pi^2}{c_{UV}}\,.
\eqlabel{ccharge}
\end{equation}
\end{itemize}

10d type IIB supergravity is invariant under $SL(2,\reals)$ duality transformation.
This duality is inherited by the consistent truncation \eqref{seff}. As a result, we
can always work in a duality frame with
\begin{equation}
C_0\equiv 0\,,
\eqlabel{c0zero}
\end{equation}
which we will do from now on.

Effective action \eqref{seff} is invariant under the 1-form gauge transformations
(with the 0-form gauge parameters $\alpha$ and $\beta$):
\nxt $A\to A+d\alpha$,
\begin{equation}
\begin{split}
&\theta\to \theta+3\alpha\,,\qquad c^\Phi\to c^\phi+q\alpha\,,\qquad b^\Phi\to b^\Phi+p\alpha\,,\\
&a\to a+k\alpha+\frac \alpha2(pc^\Phi-q b^\Phi)\,,\qquad b^\Om\to b^\Om e^{3i\alpha}\,,\qquad
c^\Om\to c^\Om e^{3i\alpha}\,.
\end{split}
\eqlabel{gaugea}
\end{equation}
Notice that $b^\Omega$, $c^\Omega$ and $\tau e^{i\theta}$ all have charge 
\begin{equation}
Q_A[b^\Om]=Q_A[c^\Om]=Q_A[\tau e^{i\theta}]=3\,,
\eqlabel{chargea}
\end{equation}
under this gauge transformation.
\nxt $a_1^J\to a_1^J+d\beta$,
\begin{equation}
\begin{split}
a\to a+2\beta\,.
\end{split}
\eqlabel{gaugeb}
\end{equation}
The gauge transformations \eqref{gaugea} and \eqref{gaugeb} can be used to eliminate
the scalars $\theta$ and $a$. Thus, we assume from now on that 
\begin{equation}
\theta\equiv 0\,,\qquad a\equiv 0\,.
\eqlabel{noatheta}
\end{equation}

Effective action \eqref{seff} is still fairly complicated --- to develop an intuition
about various parameters/fields we consider some obvious further consistent truncations.
\begin{itemize}
\item {\bf I:}\ We can set all the 3-form fluxes to zero,
\begin{equation}
b^\Phi=b^\Om=c^\Phi=c^\Om\equiv 0\,,\qquad p=q=0\,, 
\eqlabel{trunc11}
\end{equation}
as well as remove the dilaton $\phi$ and the scalar $\tau$:
\begin{equation}
\begin{split}
&S_{eff}\ \longrightarrow\ S_I\equiv \frac{1}{2\kappa_5^2}\int_{\calm_5}\ \call_I\star 1+S_{top}\\
&\call_I=R_5-\frac{28}{3} du^2-\frac 43 dv^2-\frac 83 dudv
-\frac 12 e^{\ft 83 u +\ft 83 v}\ (dA)^2-
e^{-\ft 43 u -\ft 43 v }\ (da_1^J)^2
\\&-2 e^{-8 u}\left(2a_1^J+kA\right)^2
-4e^{-\frac{20}{3}u+\frac43 v}+24e^{-\frac{14}{3}u-\frac23 v}-2k^2 e^{-\frac{32}{3}u-\frac83 v}\,.
\end{split}
\eqlabel{trunc12}
\end{equation}
\nxt Notice that in the absence of vector fields, \ie for $A=a_J\equiv 0$ (which is a consistent truncation),
for $k=2$ the scalars $u$ and $v$ can be consistently set to zero, resulting in
\begin{equation}
\call_I\bigg|_{k=2,u=v\equiv 0, A=a_i^J\equiv 0}=R_5 +12\,,
\eqlabel{trunc13}
\end{equation}
which is an effective action realizing asymptotically $AdS_5$ solution of radius $L=1$; from now on we set
\begin{equation}
k=2\,.
\eqlabel{trunc13a}
\end{equation}
\nxt We now introduce linear combinations of the vector fields
\begin{equation}
A=\frac 13 \cala+\frac 23 \calv\,,\qquad a_1^J=\frac 13 \calv-\frac 13\cala\,,
\eqlabel{trunc14}
\end{equation}
resulting in
\begin{equation}
\begin{split}
&-\frac 12 e^{\ft 83 u +\ft 83 v}\ (dA)^2-
e^{-\ft 43 u -\ft 43 v }\ (da_1^J)^2-8 e^{-8 u}\left(a_1^J+A\right)^2\\
\longrightarrow&\\
&
-\frac{1}{18} \left(
e^{\frac83 u+\frac83 v}+2 e^{-\frac43 u-\frac43 v}\right) (d\cala)^2
+\frac29 \left(
e^{-\frac43 u-\frac43 v}-e^{\frac83 u+\frac83 v}\right)
d\calv d\cala\\&-\frac19 \left(
2 e^{\frac83 u+\frac83 v}+e^{-\frac43 u-\frac43 v}
\right) (d\calv)^2-8 e^{-8 u}\ \calv^2\,.
\end{split}
\eqlabel{trunc15}
\end{equation}
We can  consistently truncate $\call_I$ removing the massive vector field $\calv$,
and the scalars $u$ and $v$:
\begin{equation}
\begin{split}
&\call_I\bigg|_{u=v\equiv 0, \calv\equiv 0}=R_5 +12-\frac 16 (d\cala)^2=R_5+12-\frac{1}{12}F_{\cala,\mu\nu}F_{\cala}^{\mu\nu} \\
&S_{top}\bigg|_{\calv\equiv 0}=\frac{1}{2\kappa_5^2}\int_{\calm_5}\ \frac{1}{27}\cdot \cala\wedge d\cala\wedge d\cala
=\frac{1}{2\kappa_5^2}\int_{\calm_5}\biggl\{\frac{1}{108}\epsilon^{\mu\nu\rho\lambda\sigma}
F_{\cala,\mu\nu}F_{\cala,\rho\lambda}\cala_\sigma\biggr\}\star 1\,,
\end{split}
\eqlabel{trunc16}
\end{equation}
reproducing minimal $\caln=2$ gauged supergravity from the consistent truncation of type IIB theory
on Sasaki-Einstein manifolds with 5-form flux \cite{Buchel:2006gb}.
The $\caln=2$ graviphoton
\begin{equation}
\cala\equiv A-2 a_1^J
\eqlabel{trunc17}
\end{equation}
is holographically dual to the R-symmetry current of the boundary gauge theory.
Turning on the boundary value for the time components of $\cala$ corresponds to
introducing a chemical potential of the Klebanov-Witten gauge theory plasma.

Notice that canonically normalizing the massive vector field $\calv$ as
$\calv\to \sqrt{\frac 32}\hat\calv$
 we identify its mass as
 \begin{equation}
\frac{m^2_{\calv}}{2}= 8 \left(\sqrt{\frac 32} \right)^2 =\frac{24}{2}\qquad \Longrightarrow\qquad m_\calv^2=24\,,
\eqlabel{trunc18}
\end{equation}
which identifies the dimension of the dual gauge theory operator as \cite{Zaffaroni:2000vh}
\begin{equation}
\dim \calo_\calv = 2+\sqrt{1+m_\calv^2}=7\,.
\eqlabel{trunc19}
\end{equation}
\item {\bf II:} Another simple consistent truncation of \eqref{seff} is obtained
turning off the vectors $\cala$, $\calv$, and removing some of the 3-form fluxes, namely
$\Re[b^\Om+i c^\Om]$,  $c^\Phi$, and setting the constant $p=0$.
The resulting truncation describes Klebanov-Strassler (KS) 
cascading gauge theory \cite{Klebanov:2000hb}, dual gravitational action for which was obtained in
\cite{Buchel:2010wp}\footnote{A truncation to the chirally symmetric sector of the KS theory
was obtained earlier in \cite{Aharony:2005zr}.}.  KS gravitational dual $S_{KS}$ of \cite{Buchel:2010wp} in 5d Einstein frame
is given by \cite{Buchel:2022zxl}:
\begin{equation}
\begin{split}
&S_{KS}=\frac{1}{16\pi G_5}\int_{\calm_5} {\rm vol}_{\calm_5}\ \biggl\{
R-\frac{40}{3} \left(\nabla f\right)^2-20 \left(\nabla w\right)^2-4 \left(\nabla \lambda\right)^2
-\frac 12\left(\nabla \Phi\right)^2\\
&-18 e^{-4 f -4w-\Phi}\biggl[e^{4\lambda} \left(\nabla h_1\right)^2+e^{-4\lambda} \left(\nabla h_3\right)^2\biggr]-36 e^{-4f -4w+\Phi} \left(\nabla h_2\right)^2-\calp_{flux}\\
&-\calp_{scalar}
\biggr\}\,,
\end{split}
\eqlabel{eaeh}
\end{equation}
where
\begin{equation}
\begin{split}
\calp_{flux}=&81 e^{-\frac{28}{3}f +4w-\Phi}(h_1-h_3)^2+162  e^{-\frac{28}{3}f +4w+\Phi}
\biggl[e^{-4\lambda}\left(h_2-\frac 19 P\right)^2+e^{4\lambda} h_2^2\biggr]\\
&+72 e^{-\frac{40}{3}f}\biggl[h_1(P-9 h_2)+9 h_2 h_3 +36\Om_0\biggr]^2\,,
\end{split}
\eqlabel{pflux}
\end{equation}
\begin{equation}
\calp_{scalar}=4 e^{-\frac{16}{3}f-12 w}-24e^{-\frac{16}{3}f-2w}\cosh(2\lambda)-\frac 92 e^{-\frac{16}{3}f+8w}
\biggl(1-\cosh(4\lambda)\biggr)\,.
\eqlabel{pscalar}
\end{equation}
Comparing \eqref{eaeh} and \eqref{seff} we identify\footnote{We indicated the dimension of the dual operators obtained
in \cite{Aharony:2005zr} and \cite{Buchel:2010wp}.} (see also section
\ref{linearized} )
\begin{equation}
\begin{split}
&u-v\equiv 5w\,,\qquad \dim[\calo_w]=6\,,\\
&4u+v\equiv 5f\,,\qquad \dim[\calo_f]=8\,,\\
&\tau\equiv 2\lambda\,,\qquad \dim[\calo_\lambda]=3\,,\\
&\Im[b^\Om-i c^\Om ]\equiv 3(h_3-h_1)-6h_2+\frac{P}{3}\,,\qquad
\dim[\calo_-]=3\,,\\
&b^\Phi\equiv 3(h_3+h_1)\,,\qquad \dim[\calo_{h_3+h_1}]=4\,,\\
&\Im[b^\Om+i c^\Om ]\equiv 3(h_3-h_1)+6h_2-\frac{P}{3}\,,\qquad \dim[\calo_{+}]=7\,,\\
&\phi\equiv \Phi\,,\qquad \dim[\calo_\Phi]=4\,;
\qquad q=-P\,,\qquad k=216 \Om_0\,.\\
\end{split}
\eqlabel{matching}
\end{equation}
The constant $P$ in \eqref{eaeh} is quantized, and is related to the number of D5 branes wrapping the 2-cycle
of the conifold base. This parameter explicitly breaks the conformal invariance of the boundary
gauge theory\footnote{By $SL(2,\reals)$ duality of type IIB 10d supergravity, parameter $p$ is related
to the number of NS5 branes wrapping the conifold base. This parameter thus explicitly breaks the
conformal invariance of the boundary gauge theory as well.}.
\end{itemize}

To summarize, the effective action $S_{KW}$ dual to Klebanov-Witten $\caln=1$ SCFT, which is relevant to study
of its thermal equilibrium states with finite $U(1)$ R-symmetry chemical potential is
a functional of a 5d metric, 2 vector fields,  6 real and 2 complex scalar fields:
\begin{equation}
S_{KW}\biggl[g_{\mu\nu}\,;\ \phi\,;\ \underbrace{u,v,\tau}_{\rm geometry}\,;\ \underbrace{b^\Om,b^\Phi}_{\rm NSNS\ 3-form\ flux}\,;\
\underbrace{c^\Om,c^\Phi}_{\rm RR\ 3-form\ flux}\,;\ \cala,\calv\biggr]=S_{eff}\bigg|_{p=0\,,q=0\,, k=2}\,,
\eqlabel{skw1}
\end{equation}
with additional  constraints \eqref{c0zero} and \eqref{noatheta}.

\subsection{Thermal charged states of the KW plasma}\label{linearized}

To consider R-charged black holes in $S_{KW}$ \eqref{skw1}, which realize the gravitational dual
to equilibrium thermal states of the KW gauge theory plasma at finite R-symmetry chemical potential,
we take the following ansatz:
\begin{equation}
\begin{split}
&ds_5^2=g_{\mu\nu}dx^\mu dx^\nu= -\hat{c}_1^2\ dt^2+\hat{c}_2^2\ d\bm{x}^2+\hat{c}_3^2\ dr^2\,,\qquad \hat{c}_i
=\hat{c}_i(r)\,,\\
&\cala=\cala_t(r)\ dt+\cala_r(r)\ dr\,,\qquad \calv=\calv_t(r)\ dt+\calv_r(r)\ dr\,,\\
&\{\phi,u,v,\tau\}=\{\phi,u,v,\tau\}(r)\,,\qquad \{b^\Om,c^\Om\}=\{b^\Om_r+i b^\Om_i,c^\Om_r+i c^\Om_i\}(r)\,,\\
&\{b^\Phi,c^\Phi\}=\{b^\Phi,c^\Phi\}(r)\,.
\end{split}
\eqlabel{backw1}
\end{equation}

Note that the equations of motion
\begin{equation}
\frac{\delta S_{KW}}{\delta \cala_r}=0\,,\qquad \frac{\delta S_{KW}}{\delta \calv_r}=0\,,
\eqlabel{gf}
\end{equation}
enforce the gauge fixing conditions \eqref{noatheta}. While not obvious at the level
of the effective action $S_{KW}$, at the level of the equations of motion within the
background ansatz \eqref{backw1},  we find two physically equivalent truncations:
\begin{equation}
\begin{split}
&{\bm{(1)}:}\qquad \cala_r=\calv_r=b_r^\Om=c_i^\Om=c^\Phi\equiv 0\,;\\
&{\bm{(2)}:}\qquad \cala_r=\calv_r=c_r^\Om=b_i^\Om=b^\Phi\equiv 0\,.\\
\end{split}
\eqlabel{backw2}
\end{equation}
From now on we work within a subtruncation $\bm{(1)}$.

It is convenient to introduce
\begin{equation}
\begin{split}
&\phi\equiv \ln g\,,\qquad \tau\equiv \ln H\,,\qquad v\equiv \ln f_1\,,\qquad u\equiv \ln f_2\,, \\
&\hat{c_1}\equiv c_1\ f_1^{1/3}f_2^{4/3}\,,\qquad \hat{c_2}\equiv c_2\ f_1^{1/3}f_2^{4/3}\,,\qquad
\hat{c_3}\equiv c_3\ f_1^{1/3}f_2^{4/3}\,.
\end{split}
\eqlabel{backw3}
\end{equation}

It is straightforward to obtain equations of motion
for the fields in \eqref{backw3} as well as $\cala_t,\calv_t,b_i^\Om,b^\Phi,c_r^\Om$ --- we will not present 
here the general equations of motion (they are too long and not illuminating).
There is an obvious solution to the equations of motion, representing the
R-symmetric phase of the KW plasma:
\begin{equation}
\begin{split}
&c_1=\frac{\alpha\sqrt{f}}{\sqrt{r}}\,,\qquad c_2=\frac{\alpha}{\sqrt{r}}\,,\qquad c_3=\frac{s}{2r\sqrt{f}}
\,,\qquad f=1-r^2\left(1+\frac{\mu^2}{9\alpha^2}\right)+\frac{\mu^2}{9\alpha^2} r^3\,,\\
&\cala_t=\mu (1-r)\,,\qquad g=s=f_1= f_2= H\equiv 1\,,\qquad \calv_t=b_i^\Om=b^\Phi=c_r^\Om\equiv 0\,,
\end{split}
\eqlabel{rnbh}
\end{equation}
where the radial coordinate $r\in (0,1)$, with $r\to 0_+$ being the asymptotic
$AdS_5$ boundary, and $r\to 1_-$ being a regular Schwarzschild horizon.
Parameter $\mu$ is a chemical potential, and $\alpha$ determines the
Hawking temperature of the black hole:
\begin{equation}
T=\frac{18\alpha^2-\mu^2}{18\pi\alpha}\,.
\eqlabel{tkw}
\end{equation}
Without the loss of generality we can always set $\alpha=1$, as long as we remember to represent
all the physical quantities as dimensionless ratios --- we will do so from now on.
Note that the chemical potential in the symmetric phase varies as
\begin{equation}
\mu\in [0,\sqrt{18})\,;
\eqlabel{murange}
\end{equation}
as $\mu\to \sqrt{18}$ the Hawking temperature vanishes, and the black hole becomes extremal.

Next, we move to the discussion of the generic fluctuations about the
symmetric background \eqref{rnbh}. Introducing
\begin{equation}
\begin{split}
&f=\biggl(1-r^2\left(1+\frac{\mu^2}{9}\right)+\frac{\mu^2}{9} r^3\biggr)(1+\delta f)\,,\qquad s=1+\delta s\,,\qquad
\cala_t=\mu (1-r)(1+\delta \cala)\,, \\
&f_1=1+\delta f_1\,,\qquad f_2=1+\delta f_2\,,\qquad H=1+\delta H\,,\qquad \calv_t=\delta \calv\,,\qquad b_i^\Om
=\delta b_i^\Om\,,\\
&b^\Phi=\delta b^\Phi\,,\qquad \qquad c_r^\Om=\delta c_r^\Om\,,\qquad g=1+\delta g\,,
\end{split}
\eqlabel{fluc1}
\end{equation}
to linear order in $\delta$-fluctuations we find the following decoupled sets\footnote{We indicated
the scaling dimensions of the dual operators $\dim\calo_{..}\equiv \Delta_{\cdots}$, obtained from the asymptotic fall-off
of the corresponding fluctuation near the $AdS_5$ boundary, \ie as $r\to 0_+$.}: 
\begin{itemize}
\item (A):\ $\Delta_{b^\Phi}=4$,
\begin{equation}
0=(\delta b^\Phi)''+ \frac{2 \mu^2 r^3-\mu^2 r^2-9 r^2-9}{r (\mu^2 r^3-\mu^2 r^2-9 r^2+9)}\ (\delta b^\Phi)'\,;
\eqlabel{casea}
\end{equation}
\item (B):\ $\Delta_{-}=3$, $\qquad \calo_-\ \longleftrightarrow\ \Im[b^\Om-ic^\Om]\ \longleftrightarrow\ \delta b_i^\Om-
\delta c_r^\Om \equiv \delta f_-$,
\begin{equation}
0=(\delta f_-)''+\frac{2 \mu^2 r^3-\mu^2 r^2-9 r^2-9}{r (r-1) (\mu^2 r^2-9 r-9)} (\delta f_-)'+
\frac{27(4 \mu^2 r^2-3 \mu^2 r-9 r-9)}{4(r-1) (\mu^2 r^2-9 r-9)^2 r^2} \delta f_-\,;
\eqlabel{caseb}
\end{equation}
\item (C):\ $\Delta_{+}=7$, $\qquad \calo_+\ \longleftrightarrow\ \Im[b^\Om+ic^\Om]\ \longleftrightarrow\ \delta b_i^\Om+
\delta c_r^\Om \equiv \delta f_+$,
\begin{equation}
0=(\delta f_+)''+\frac{2 \mu^2 r^3-\mu^2 r^2-9 r^2-9}{r (r-1) (\mu^2 r^2-9 r-9)} (\delta f_+)'
- \frac{27(4 \mu^2 r^2+3 \mu^2 r-63 r-63)}{4(r-1) (\mu^2 r^2-9 r-9)^2 r^2}\delta f_+\,;
\eqlabel{casec}
\end{equation}
\item (D):\ $\Delta_{g}=4$,
\begin{equation}
0=(\delta g)''+ \frac{2 \mu^2 r^3-\mu^2 r^2-9 r^2-9}{r (\mu^2 r^3-\mu^2 r^2-9 r^2+9)}\ (\delta g)'\,;
\eqlabel{cased}
\end{equation}
\item (E):\ $\Delta_{H}=3$,
\begin{equation}
0=(\delta H)''+\frac{2 \mu^2 r^3-\mu^2 r^2-9 r^2-9}{r (r-1) (\mu^2 r^2-9 r-9)} (\delta H)'+
\frac{27(4 \mu^2 r^2-3 \mu^2 r-9 r-9)}{4(r-1) (\mu^2 r^2-9 r-9)^2 r^2} \delta H\,;
\eqlabel{casee}
\end{equation}
\item (F):\ the coupled set\footnote{Decoupling happens when $\mu=0$.}
of $\Delta_{u-v}=6\ \longleftrightarrow\ \delta f_2-\delta f_1$,
$\Delta_{4u +v}=8 \ \longleftrightarrow\ 4\delta f_2+\delta f_1$ (see \eqref{backw3}), and
$\Delta_{\calv}=7$ (see \eqref{trunc19}),
\begin{equation}
\begin{split}
&0=(\delta f_1)''+\frac{2 \mu^2 r^3-\mu^2 r^2-9 r^2-9}{r (r-1) (\mu^2 r^2-9 r-9)} (\delta f_1)'-\frac{3 \mu r}{(r-1) (\mu^2 r^2-9 r-9)}
(\delta\calv)'\\
&+\frac{2 (\mu^2 r^3-18)}{r^2 (r-1) (\mu^2 r^2-9 r-9)} (\delta f_1+\delta f_2)\,,\\
&0=(\delta f_2)''+\frac{2 \mu^2 r^3-\mu^2 r^2-9 r^2-9}{r (r-1) (\mu^2 r^2-9 r-9)} (\delta f_2)'-\frac{9(\delta f_1+7 \delta f_2) }{r^2 (r-1) (\mu^2 r^2-9 r-9)}
\,,\\
&0=(\delta\calv)''-\frac{54}{r^2 (\mu^2 r^3-\mu^2 r^2-9 r^2+9)} \delta\calv-\frac{4\mu}{3}((\delta f_1)'+(\delta f_2)')\,;
\end{split}
\eqlabel{casef}
\end{equation}
\item (G): the remaining fluctuations are being sourced by the set $(F)$,
\begin{equation}
\begin{split}
&0=(\delta\cala)''+\frac{2}{r-1}(\delta\cala)' +\frac{(\delta f_1)'-3 (\delta s)'+4 (\delta f_2)'}{r-1}\,,
\\
&0=(\delta f)'+\frac{\mu^2 r^3-18}{r (r-1) (\mu^2 r^2-9 r-9)} \delta f
-\frac{2 r^2 \mu^2 (\delta\cala)'}{\mu^2 r^2-9 r-9}-\frac{2 r^2 \mu^2\delta\cala}{(r-1) (\mu^2 r^2-9 r-9)} \\
&+\frac{8(\mu^2 r^3-2 \mu^2 r^2-18 r^2+36)}{3(r-1) (\mu^2 r^2-9 r-9)} \left((\delta f_2)'+\frac14 (\delta f_1)'\right)
+\frac{36\delta s}{r (r-1) (\mu^2 r^2-9 r-9)} \\&+\frac{2(\mu^2 r^3+18) (\delta f_1+4 \delta f_2)}{3r (r-1) (\mu^2 r^2-9 r-9)}\,,\\
&0=(\delta s)'+\frac{2 \mu r^2  (\delta\calv)'}{(r-1) (\mu^2 r^2-9 r-9)}
-\frac{((\delta f_1)'+4 (\delta f_2)')(\mu^2 r^3+\mu^2 r^2+9 r^2-45)}{3(r-1) (\mu^2 r^2-9 r-9)} \\
&-\frac{4((\delta f_1 + \delta f_2 )r^3\mu^2-36 (\delta f_1+4 \delta f_2))}{3r (r-1) (\mu^2 r^2-9 r-9)}\,.
\end{split}
\eqlabel{caseg}
\end{equation}
\end{itemize}

We need to analyze the stability of the fluctuation sets as the chemical potential varies within \eqref{murange}.
Some sets can be analyzed analytically; the other ones numerically. For example, from \eqref{cased}, we find
\begin{equation}
\delta g=G_1\left\{\
\frac{\ln(1-r)}{\mu^2-18}-\frac{\ln(9+9 r-\mu^2 r^2)}{2(\mu^2-18)}+\frac{9\arctanh\left(\frac{2\mu^2 r-9}{3\sqrt{4\mu^2+9}}\right)}
{(\mu^2-18)\sqrt{4\mu^2+9}}
\ \right\}+G_2\,,
\eqlabel{solved}
\end{equation}
where $G_1$ and $G_2$ are the integration constants. Normalizability at the boundary requires $G_2=0$; the mode is singular at the 
horizon for all values of $\mu$ as in \eqref{murange} --- it never develops an instability. 

We find that the only unstable fluctuations are (B) and (E), associated with the condensation of the dimension-3 operators,
see \eqref{caseb} and \eqref{casee}. Recall that the fluctuation $\delta f_-$  (case (B)) is associated with the fluctuation
of the 3-form fluxes --- $\Im(b^\Om-ic^\Om)=\colorbox{yellow}{$\Im(b^\Om)$}-\colorbox{yellow}{$\Re(c^\Om)$}$ --- these are the modes
that would lead to the spontaneous fluxification of the conifold black holes. On the other hand, the
fluctuation $\delta H$  (case (E)) is associated with the fluctuation
of the scalar warp factors, deforming the conifold base   ---  \colorbox{yellow}{$\tau e^{i\theta}$}  --- this is the  mode
that would lead to the spontaneous scalarization of the conifold black holes. Notice that the equations
describing the fluctuations $\delta f_- $ and $\delta H$ are identical. This is not a surprise since both modes
are dual to chiral primary operators of the KW gauge theory of the (properly normalized) R-charge 2, \cite{Ceresole:1999zs,Cassani:2010na}:
\begin{equation}
\Im(b^\Om-i c^\Om)\ \Longleftrightarrow\ \rm{Tr}\left(W_1^2+W_2^2\right)\,,\qquad \tau e^{i\theta} \Longleftrightarrow\ \rm{Tr}\left(W_1^2-W_2^2\right)\,,
\eqlabel{cp}
\end{equation}
where $W_i$ are the gauge superfields of the KW quiver gauge theory.
To see that the properly normalized R-charge of $\delta f_-$ and $\delta H$ modes is indeed 2, note that 
for the boundary gauge theory superpotential to have a standard R-charge 2, the  bulk gauge field $\hat F$ dual to the R-symmetry current
should be normalized as $-\frac 13 \hat{F}_{\mu\nu}\hat{F}^{\mu\nu}$ \cite{Berenstein:2002ke}. Instead, our gauge field $\cala$ is normalized as
in \eqref{trunc16}, \ie $\cala=2 \hat{\cala}$. Under the gauge transformation $A\to A+d\alpha$,
equivalently (see \eqref{trunc14})
\begin{equation}
\cala\to \cala+3\ d(\alpha)\qquad \Longleftrightarrow\qquad \hat{\cala}\to \hat{\cala}+\frac32\ d(\alpha)\equiv \hat{\cala}+d(\hat{\alpha})\,,
\end{equation}
$f_-$ and $\tau e^{i\theta}$ transform as
charge 3 objects, see \eqref{chargea},
\begin{equation}
\{f_-\,,\, \tau e^{i\theta}\}\ \longrightarrow\ \{f_-\,,\, \tau e^{i\theta}\} e^{3i \alpha}=\{f_-\,,\, \tau e^{i\theta}\} e^{3i\cdot \frac 23 \hat{\alpha}}
=\{f_-\,,\, \tau e^{i\theta}\} e^{2i\hat{\alpha}}\,.
\eqlabel{propcha}
\end{equation}
The lowest components of the superfields form fermion bilinears and
condense in the symmetric phase of the charge KW gauge theory plasma at critical value of the chemical potential (see appendix \ref{mode3})
\begin{equation}
\frac{T}{\mu}\bigg|_{crit}=0.030338(3)\,.
\eqlabel{tmuc}
\end{equation}
Note that our onset of the instability \eqref{tmuc} is precisely twice the value for the identical chiral primary instability
reported in \cite{Gubser:2009qm} --- this discrepancy is simply a reflection of the fact that the authors of \cite{Gubser:2009qm}
used properly normalized bulk gauge field, and so $\mu|_{here}=2\hat{\mu}|_{there}$.

Remarkably, there is a fully nonlinear further consistent truncation of $\bm(1)$ of \eqref{backw1} that retains either
one\footnote{There is also a consistent truncation when both $\ln H\ne 0$ and $K\ne 0$, but in this case all
the modes in \eqref{zeros} are excited as well.} of the  unstable modes $H$ or $f_-\equiv \Im[b^\Om-i c^\Om]\equiv -K$,
while setting\footnote{This consistent truncation
is 'gauging' of the novel truncation on the conifold identified in \cite{Buchel:2024hdd}.}
\begin{equation}
\underbrace{\ln g}_{\rm dilaton}\equiv 0\,,\qquad \underbrace{b^\Phi}_{\Delta_{b^\Phi}=4}\equiv 0\,,\qquad \underbrace{\Im[b^\Om+i c^\Om]}_{\Delta_{f_+}=7}\equiv 0\,.
\eqlabel{zeros}
\end{equation}
The equations of motion for the 5d metric factors $c_i$, the time-components of the bulk vectors $\cala_t$ and $\calv_t$,
the scalars $f_1,f_2,H$ (see \eqref{backw3}), and the 3-form flux mode $K$ are collected in appendix \ref{kweoms}. 
There are two distinct truncations of these equations of motions,
\nxt $\bm{(1S)}$: $K=0$. Here, the corresponding black hole develops {\it scalar hair}.
The massive vector field $\calv$ is excited.
\nxt $\bm{(1F)}$: $\ln H=0$. Here, the corresponding black hole develops {\it flux hair}.
A detailed but highly nontrivial analysis of the equations of motion \eqref{kwpap1}-\eqref{kwpapc} shows that in this case
there is an additional truncation of the equations of motion, eliminating the massive vector:
\begin{equation}
\calv_t\equiv 0\,,\qquad f_1\equiv F\,,\qquad f_2\equiv \frac1F\,,\qquad F^4(K^2-1)+4\equiv 0 \,,
\eqlabel{fintrunc}
\end{equation}
\ie the scalar warp factors of the conifold base, namely $f_1$ and $f_2$, are algebraically determined
by the flux $K$. 

Parameterizing the 5d metric  as in \eqref{rnbh},
\begin{equation}
c_1=\frac{\sqrt f}{\sqrt r}\,,\qquad c_2=\frac{1}{\sqrt r}\,,\qquad c_3=\frac{s}{2r\sqrt{f}}\,,
\eqlabel{5dmetf}
\end{equation}
the equations of motion describing black holes with flux/scalar hair are solved subject to the following
asymptotics\footnote{Again, this has to be understood subject to constraint $K\cdot \ln H=0$, \ie
either $K\equiv 0$ or $H\equiv 1$}:
\nxt in the UV, \ie as $r\to 0_+$,
\begin{equation}
\begin{split}
&f=1+r^2 f_4+\frac{a_2^2}{9} r^3+\left(\frac14 H_3^2 \mu^2+\frac{1}{32} \mu^2 K_3^2\right) r^4+\calo(r^5)\,,
\end{split}
\eqlabel{uvgen1}
\end{equation}
\begin{equation}
\begin{split}
&s=1+\left(-\frac35 H_3^2+\frac{3}{16} K_3^2\right) r^3+\left(
\frac{929}{120} H_3^2 \mu^2-\frac{61}{192} \mu^2 K_3^2-15 f_{1,8}+\frac38 H_3^2 \mu^2 \ln r\right) r^4\\
&\qquad +\calo(r^5\ln r)\,,
\end{split}
\eqlabel{uvgen2}
\end{equation}
\begin{equation}
\begin{split}
&\cala_t=\mu+a_2 r+\left(
\frac18 \mu K_3^2+H_3^2 \mu\right) r^3+\left(-\frac{1}{64} \mu^3 K_3^2+\frac25 a_2 H_3^2+\frac18 a_2 K_3^2-\frac18 H_3^2 \mu^3\right) r^4\\&
\qquad +\calo(r^5\ln r)\,,
\end{split}
\eqlabel{uvgen3}
\end{equation}
\begin{equation}
\begin{split}
&\calv_t=\left(v_6+\frac65 H_3^2 \mu \ln r\right) r^3+\biggl(\frac{1}{10} a_2 H_3^2+\frac{1}{32} a_2 K_3^2-\frac12
a_2 f_{1,6}-\frac14 H_3^2 \mu^3\\&\qquad +\frac65 a_2 H_3^2 \ln r\biggr) r^4 +\calo(r^5\ln r)\,,
\end{split}
\eqlabel{uvgen4}
\end{equation}
\begin{equation}
\begin{split}
&f_1=1+r^3 \left(f_{1,6}-\frac{12}{5} H_3^2 \ln r\right)+r^4 \left(f_{1,8}-\frac{1}{40} H_3^2 \mu^2 \ln r\right)+\calo(r^5\ln r)\,,
\end{split}
\eqlabel{uvgen5}
\end{equation}
\begin{equation}
\begin{split}
&f_2=1+r^3 \left(-\frac{3}{20} H_3^2-\frac{3}{64} K_3^2-\frac14 f_{1,6}+\frac35 H_3^2 \ln r\right)+r^4 \biggl(
-\frac35 H_3^2 \mu^2+\frac{1}{32} \mu^2 K_3^2+f_{1,8}\\&\qquad -\frac{1}{40} H_3^2 \mu^2 \ln r\biggr)+\calo(r^5\ln r)\,,
\end{split}
\eqlabel{uvgen6}
\end{equation}
\begin{equation}
\begin{split}
&H=1+H_3 r^{3/2}-\frac18 H_3 \mu^2 r^{5/2}+\frac12 H_3^2 r^3+\biggl(
\frac{1}{192} H_3 \mu^4-\frac{1}{12} H_3 a_2 \mu-\frac{3}{8} f_4 H_3\biggr) r^{7/2}\\
&\qquad -\frac18 H_3^2 \mu^2 r^4+\calo(r^{9/2}\ln r)\,,
\end{split}
\eqlabel{uvgen7}
\end{equation}
\begin{equation}
\begin{split}
&K=K_3 r^{3/2}-\frac18 K_3 \mu^2 r^{5/2}+\left(\frac{1}{192} \mu^4 K_3-\frac{1}{12} a_2 \mu K_3-\frac38 K_3 f_4\right) r^{7/2}+\calo(r^{9/2}\ln r)\,,
\end{split}
\eqlabel{uvgen8}
\end{equation}
specified by 
\begin{equation}
\biggl\{\
a_2\,,\, H_3\,,\, K_3\,,\, f_4\,,\, f_{1,6}\,,\, v_6\,,\, f_{1,8}
\
\biggr\}\,,
\eqlabel{uvpar}
\end{equation}
as functions of a chemical potential $\mu$; 
\nxt in the IR, \ie as $y\equiv 1-r\to 0_+$,
\begin{equation}
\begin{split}
&f_1=f_{1,0}^h+\calo(y)\,,\qquad f_2=f_{2,0}^h+\calo(y)\,,\qquad K=K_{0}^h+\calo(y)\,,\qquad H=H_{0}^h+\calo(y)\,,\\
&s=s^h_0+\calo(y)\,,\qquad \cala_t=a_1^h\ y+\calo(y^2)\,,\qquad \calv_t=v_1^h\ y+\calo(y^2)\,,\\
&f=\biggl(\frac{9(K^h_0)^4 (s^h_0)^2}{32(f^h_{2,0})^8 (f^h_{1,0})^2}+
\frac{3(s^h_0)^2  (K^h_0)^2}{64(f^h_{2,0})^8 (f^h_{1,0})^2 (H^h_0)^4}\biggl((3 (H^h_0)^8 (f^h_{2,0})^4+6 (H^h_0)^4 (f^h_{2,0})^4\\
&-32 (H^h_0)^4+3 (f^h_{2,0})^4\biggr)
-\frac{(a^h_1-v^h_1)^2}{9(f^h_{2,0})^4 (f^h_{1,0})^2}+\frac{2 (s^h_0)^2}{(f^h_{2,0})^8 (f^h_{1,0})^2}\biggr)\ y+\calo(y^2)\,,
\end{split}
\eqlabel{irass}
\end{equation}
specified by 
\begin{equation}
\biggl\{\
s^h_0\,,\, f_{1,0}^h\,,\, f^h_{2,0}\,,\, H^h_0\,,\, K^{h}_0\,,\, a_1^h\,,\, v_1^h
\
\biggr\}\,,
\eqlabel{irpar}
\end{equation}
again, as functions of a chemical potential $\mu$.

\begin{figure}[ht]
\begin{center}
\psfrag{r}[tt][][1.0][0]{{$r$}}
\psfrag{f}[bb][][1.0][0]{{$f(r)$}}
\psfrag{s}[tt][][1.0][0]{{$s(r)$}}
\includegraphics[width=3in]{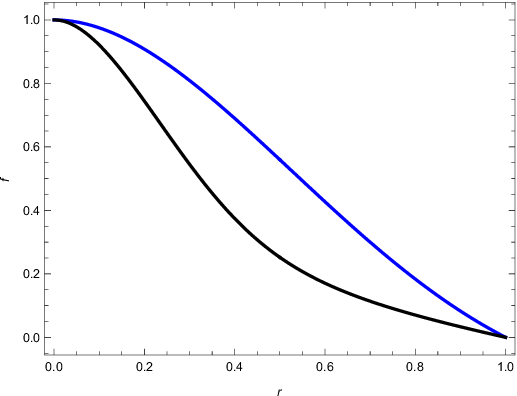}
\includegraphics[width=3in]{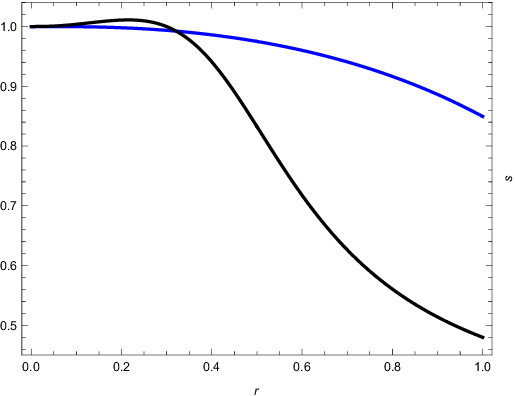}
\end{center}
  \caption{Sample of scalarized (in blue) and fluxified (in black) conifold black hole metric warp factors 
  $f$ and $s$ (see \eqref{5dmetf}) for select values of $\frac T\mu$, as a function of a radial coordinate
  $r$. $r=0$ is the asymptotic $AdS_5$ boundary, and $r=1$ is a regular black hole horizon. 
} \label{plotfs}
\end{figure}

\psfrag{y}[tt][][1.0][0]{{$\hat\cals/\hat\rho$}}
\begin{figure}[ht]
\begin{center}
\psfrag{r}[tt][][1.0][0]{{$r$}}
\psfrag{a}[bb][][1.0][0]{{$\cala_t(r)$}}
\psfrag{c}[tt][][1.0][0]{{$\ln H(r)\ {\rm or}\ K(r)$}}
\includegraphics[width=3in]{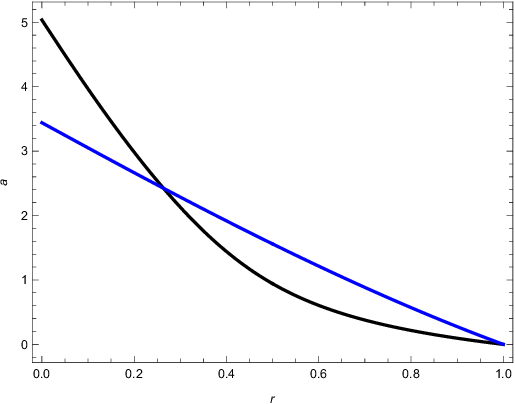}
\includegraphics[width=3in]{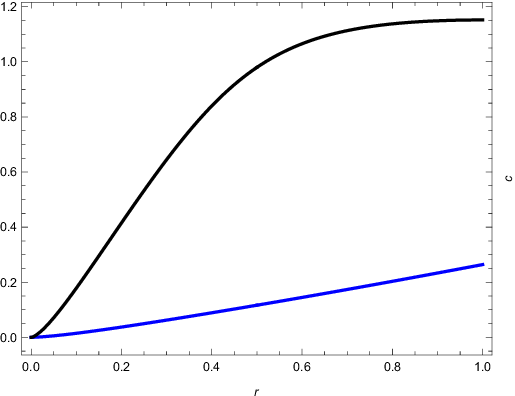}
\end{center}
  \caption{Sample of profiles of scalarized (in blue) and fluxified (in black) conifold black hole gauge field $\cala_t$ 
  and the scalar condensates  ($\ln H$ or $K$)  for select values of $\frac T\mu$, as a function of a radial coordinate
  $r$. $r=0$ is the asymptotic $AdS_5$ boundary, and $r=1$ is a regular black hole horizon. 
} \label{plotc}
\end{figure}

In practice, we solve equations of motion for $f,s,\cala_t,\calv_t,f_1,f_2$ and $K$ or $H$ (depending whether we look
for a flux hair black hole --- $\bm{(1F)}$ truncation --- or  a scalar  hair black hole --- $\bm{(1S)}$ truncation)
using the shooting method codes adopted from \cite{Aharony:2007vg}.
In figs.~\ref{plotfs} and \ref{plotc} we plot a sample of the numerically obtained profiles
for $f,s,\cala_t$ and $\ln H$ or $K$
for the scalarized black holes (in blue) and the fluxified black holes (in black), as a function of the
radial coordinate $r$. The ``numerical shooting'' is performed from boundary ($r=0$)  and the horizon ($r=1$),
with the matching at $r=\frac 12$ achieved with an accuracy of $\sim 10^{-30}$ or better.
The scalarized conifold black hole profiles are given for $\frac T\mu \approx 1.3\ \frac {T_{crit}}{\mu }$
(see \eqref{tmucin}), and the fluxified
conifold black hole profiles are given for $\frac T\mu \approx 0.7\ \frac {T_{crit}}{\mu }$.

We keep $\calv_t$ massive vector in $\bm{(1F)}$, even though we claim it is decoupled \eqref{fintrunc}.
As a check on the accuracy of the numerics we find that in this truncation $\{v_6, v^h_1\} \sim 10^{-15}-10^{-13}$.
Additional check on numerics comes from the algebraic equation in \eqref{fintrunc},
which fixes
\begin{equation}
f_{1,6}=\frac{1}{16} H_3^2\,,\qquad f_{1,8}=-\frac{1}{64} H_3^2\mu^2\,,
\end{equation}
which we find to be satisfied to fractional accuracy of $\sim 10^{-14}$ or better.

Once hairy black hole solutions are constructed, we can use the holographic
renormalization\footnote{In this model a simple holographic renormalization of
\cite{Balasubramanian:1999re}
is enough.}
to extract their thermodynamic properties:
\begin{equation}
\begin{split}
&2\pi T=\frac{9(K^h_0)^4 s^h_0}{32(f^h_{2,0})^8 (f^h_{1,0})^2}+
\frac{3s^h_0  (K^h_0)^2}{64(f^h_{2,0})^8 (f^h_{1,0})^2 (H^h_0)^4}\biggl((3 (H^h_0)^8 (f^h_{2,0})^4+6 (H^h_0)^4 (f^h_{2,0})^4\\
&-32 (H^h_0)^4+3 (f^h_{2,0})^4\biggr)
-\frac{(a^h_1-v^h_1)^2}{9(f^h_{2,0})^4 (f^h_{1,0})^2s^h_0}+\frac{2 s^h_0}{(f^h_{2,0})^8 (f^h_{1,0})^2}\,,\\
&\hat{\cals}\equiv 2\kappa_5^2 \cals =4\pi f^h_{1,0} (f^h_{2,0})^4\,,\qquad  \hat\Omega\equiv 2\kappa_5^2 \Omega =\frac 23 \mu a_2 -3 f_4 -T\hat\cals\,,\\
&\hat\cale\equiv 2\kappa_5^2 \cale = -3\hat\Omega\,,\qquad \hat\rho\equiv  2\kappa_5^2 \rho =\frac 1\mu \left(
\hat\cale-T\hat\cals-\hat\Omega\right)\,,
\end{split}
\eqlabel{thermo}
\end{equation}
where $T$ is the temperature,  $\cals$ is the entropy density, $\Omega$ is the Gibbs free energy density,
$\cale$ is the energy density and $\rho$ is the R-symmetry charge density. 
The  thermodynamic phases  of the charged conifold black holes are presented in fig.~\ref{figure2}.

\begin{figure}[ht]
\begin{center}
\psfrag{z}[tt][][1.0][0]{{$T/\mu$}}
\psfrag{x}[bb][][1.0][0]{{$1/\hat\cals\cdot \del\hat\Omega/\del T+1$}}
\psfrag{y}[tt][][1.0][0]{{$1/\hat\cals\cdot \del\hat\Omega/\del T+1$}}
\includegraphics[width=3in]{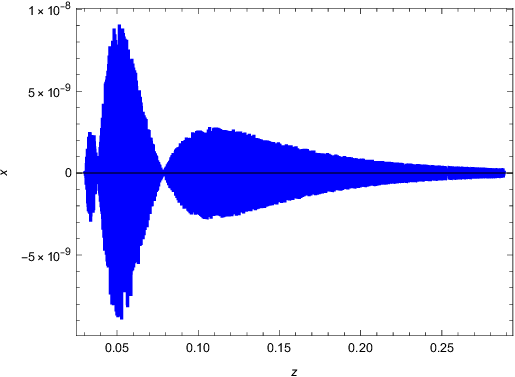}
\includegraphics[width=3in]{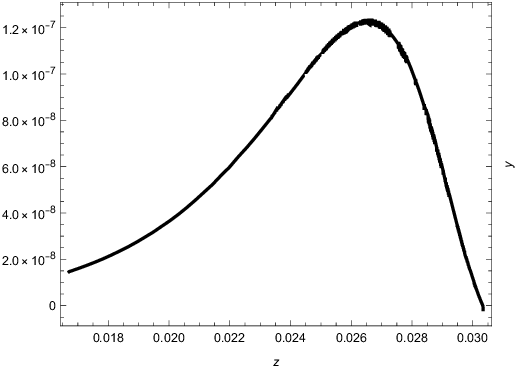}
\end{center}
  \caption{Left panel: numerical test of the first law of thermodynamics \eqref{1stlaw}
  for the scalarized conifold black holes
  (truncation $\bm{(1S)}$). Right panel: numerical test of the first law of thermodynamics \eqref{1stlaw}
  for the fluxified conifold black holes
  (truncation $\bm{(1F)}$). 
} \label{figure3}
\end{figure}

Given the R-symmetric solution \eqref{rnbh}, we identify
\begin{equation}
\begin{split}
&2\pi T=2\alpha-\frac{\mu^2}{9\alpha}\,,\qquad \hat\cals=4\pi\alpha^3\,,\qquad \hat\Omega=-\alpha^4-\frac 19\alpha^2\mu^2\,,\\
&\hat\cale=3\alpha^4+\frac 13\alpha^2\mu^2\,,\qquad \hat\rho=\frac 23\mu\alpha^2\,,
\end{split}
\eqlabel{symthermo}
\end{equation}
where we restored $\alpha$. Given \eqref{symthermo}, we can verify the fundamental thermodynamic relation
\begin{equation}
d\hat\Omega=-\hat\cals\ dT-\hat\rho\ d\mu\,.
\eqlabel{stlaw}
\end{equation}
In phases of the model with flux/scalar hair black holes, \eqref{stlaw} provides an important check on numerics.
In fig.~\ref{figure3} we present the check of the first law of thermodynamics, namely
\begin{equation}
0=\frac{\mu^3}{\hat\cals}\cdot \left(\frac{\del \frac{\hat\Omega}{\mu^4}}{\del \frac T\mu}\right)\bigg|_{\mu={\rm const}} +1\,.
\eqlabel{1stlaw}
\end{equation}

\section*{Acknowledgments}
Research at Perimeter
Institute is supported by the Government of Canada through Industry
Canada and by the Province of Ontario through the Ministry of
Research \& Innovation. This work was further supported by
NSERC through the Discovery Grants program.

\appendix
\section{Onset of instability of $\delta f_-$ and $\delta H$ fluctuations}\label{mode3}

The easiest way to identify the onset of the instability for a gravitational mode is to turn on
its source term; the instability is then signalled by the divergence of its normalizable
component\footnote{See for example section 5 of \cite{Buchel:2019pjb}.}.  
We focus on $\delta H$, as $\delta f_-$ mode is described identically.

\begin{figure}[ht]
\begin{center}
\psfrag{z}[tt][][1.0][0]{{$\mu^2$}}
\psfrag{x}[bb][][1.0][0]{{$1/h_{3,0}$}}
\psfrag{y}[tt][][1.0][0]{{$1/h^h_0$}}
\includegraphics[width=3in]{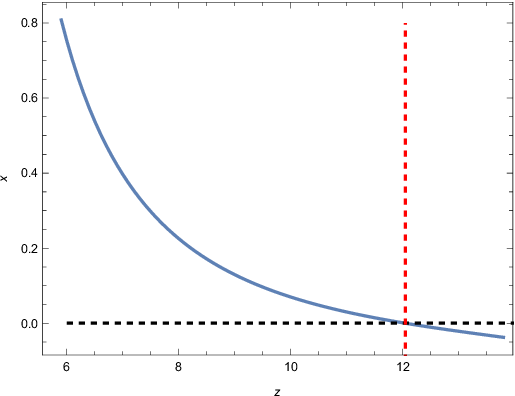}
\includegraphics[width=3in]{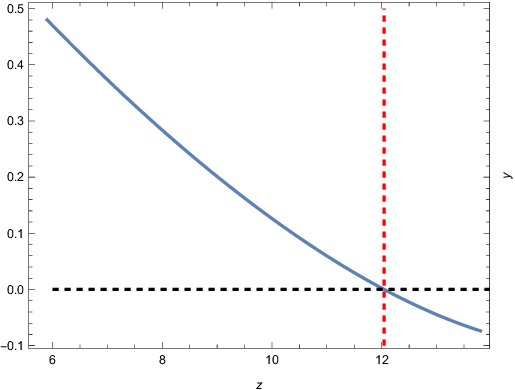}
\end{center}
  \caption{At the onset of the instability, the normalizable component $h_{3,0}$ of the bulk mode $\delta H$ diverges (the left panel).
  Likewise, the value of the fluctuation at the black hole horizon diverges (the right panel). The vertical dashed red lines
  indicate the critical value of the chemical potential $\mu_c$
} \label{figure1}
\end{figure}

The relevant equation is \eqref{casee}.
\nxt In the UV, \ie as $r\to 0_+$, we find
\begin{equation}
\delta H=r^{1/2}\left(\colorbox{red}{1}+\left(\colorbox{yellow}{$h_{3,0}$}-\frac{\mu^2}{4}\ln r\right) r+\calo(r^2\ln r)
\right)\,,
\eqlabel{uvh}
\end{equation}
where we highlighted the source term $\colorbox{red}{1}$ (which can be set to 1 as the equation is linear), and the
normalizable coefficient $\colorbox{yellow}{$h_{3,0}$}$.
\nxt Regularity at the black hole horizon identifies the asymptotic expansion $y\equiv 1-r\to 0_+$ as
\begin{equation}
\delta H=h^h_0\biggl(1+\frac{27}{4(\mu^2-18)}\ y+\calo(y^2)\biggr)\,.
\eqlabel{irh}
\end{equation}
We use the shooting method code developed in \cite{Aharony:2007vg} to
determine $\{\frac {1}{h_3}, \frac{1}{h^h_0}\}$ as we vary $\mu$ \eqref{murange}.
The results of the numerics are presented in fig.~\ref{figure1}.
Both coefficients vanish at
\begin{equation}
\mu_{crit}^2= 12.045(7)\,,
\eqlabel{critmu}
\end{equation}
corresponding to $\frac T\mu$ in \eqref{tmuc}.

\section{Equations of motion}\label{kweoms}

We would like to treat consistent truncations $\bf (1S)$ and $\bf (1F)$ simultaneously,
but it has to be remembered that one always must set either $K\equiv 0$ or $H\equiv 1$
in the equations below. When both of these modes are non-vanishing,  \ie $K\cdot \ln H\ne 0$,
additional fields, specifically the ones in \eqref{zeros}, will be sourced.  
 
There are 8 second-order coupled equations of motion for $\{c_{1,2,3},f_{1,2},H,K,\cala_t,\calv_t\}$:
\begin{equation}
\begin{split}
&0=(\cala_t)''+\biggl(
-\frac{c_1'}{c_1}-\frac{c_3'}{c_3}+\frac{3c_2'}{c_2}+\frac{f_1'}{3f_1}+\frac{4f_2'}{3f_2}
\biggr) (\cala_t)'
+\left(
\frac{8f_2'}{3f_2}+\frac{8f_1'}{3f_1}
\right) (\calv_t)'
\\&-\frac{3 c_3^2 (2 f_2^4+K^2) (\cala_t+2 \calv_t)H^4}{4f_2^4 f_1^2}
-\frac{3c_3^2 K^4 (\cala_t+2 \calv_t)}{4f_2^8 f_1^2}
-\frac{c_3^2 K^2}{4 f_2^8 H^4 f_1^2} \biggl(
(2 H^4 (4 f_1^4 f_2^4+3 f_2^4\\&-4)+3 f_2^4) \cala_t+2 \calv_t (2 H^4 (4 f_1^4 f_2^4+3 f_2^4-10)+3 f_2^4)\biggr)
+\frac{c_3^2}{2 f_2^8 H^4 f_1^2} \biggl(
3 f_2^8 (2 H^4-1) \cala_t\\&+2 \calv_t (2 H^4 (8 f_1^4 f_2^4+3 f_2^8-8)-3 f_2^8)
\biggr)\,;
\end{split}
\eqlabel{kwpap1}
\end{equation}
\begin{equation}
\begin{split}
&0=(\calv_t)''+\biggl(
\frac{8f_2'}{3f_2}+\frac{5f_1'}{3f_1}-\frac{c_1'}{c_1}-\frac{c_3'}{c_3}+\frac{3c_2'}{c_2}
\biggr) (\calv_t)'
+\left(
\frac{4f_2'}{3f_2}+\frac{4f_1'}{3f_1}
\right) (\cala_t)'\\
&-\frac{3c_3^2 H^4(2 f_2^4+K^2) (\cala_t+2 \calv_t)}{4f_2^4 f_1^2}
-\frac{3c_3^2K^4 (\cala_t+2 \calv_t)}{4f_2^8 f_1^2}
+\frac{c_3^2 K^2}{4 H^4 f_2^8 f_1^2} \biggl(
(2 H^4 (2 f_1^4 f_2^4-3 f_2^4\\&+4)-3 f_2^4) \cala_t+2 \calv_t (2 H^4 (2 f_1^4 f_2^4-3 f_2^4+10)-3 f_2^4)\biggr)
-\frac{c_3^2}{2 H^4 f_2^8 f_1^2} \biggl(
-3 f_2^8 (2 H^4-1) \cala_t\\&+2 \calv_t (2 H^4 (4 f_1^4 f_2^4-3 f_2^8+8)+3 f_2^8)
\biggr)\,;
\end{split}
\eqlabel{kwpap2}
\end{equation}
\begin{equation}
\begin{split}
&0=c_1''-\frac{3c_1(c_2')^2}{8c_2^2} +\biggl(
-\frac{c_3'}{c_3}+\frac{7f_2'}{2f_2}+\frac{21c_2'}{8c_2}+\frac{7f_1'}{8f_1}
\biggr) c_1'
-\frac{3c_1 c_2'}{8c_2} \biggl(\frac{f_1'}{f_1}+\frac{4f_2'}{f_2}\biggr)
\\&-\frac{3c_1 (H^4+1)^2(K')^2}{128f_2^4 H^4} +\frac{c_1(H')^2}{4H^2}-\frac{(17 f_1^4 f_2^4+18)(\cala_t')^2}{288f_2^4 f_1^2 c_1}
-\frac{(34 f_1^4 f_2^4+9)(\calv_t')^2}{144f_2^4 f_1^2 c_1} \\&
-\frac{\calv_t' \cala_t' (17 f_1^4 f_2^4-9)}{72f_2^4 f_1^2 c_1}
-\frac{3c_1(f_2')^2}{4f_2^2}
-\frac{f_2' f_1' c_1}{2f_1 f_2}
-\frac{c_3^2K^4}{128f_2^8 f_1^2 c_1} \biggl(7 f_1^2 \cala_t^2+28 f_1^2 \cala_t \calv_t+28 f_1^2 \calv_t^2\\
&+81 c_1^2\biggr)
-\frac{K^2 c_3^2}{128 H^4 f_2^8 f_1^2 c_1} \biggl(
f_2^4 (11 f_1^2 \cala_t^2+44 f_1^2 \cala_t \calv_t+44 f_1^2 \calv_t^2+45 c_1^2) H^8+(22 f_1^2 f_2^4 \cala_t^2\\&
+88 f_1^2 f_2^4 \cala_t \calv_t+88 f_1^2 f_2^4 \calv_t^2+90 f_2^4 c_1^2-112 f_1^2 \cala_t \calv_t-224 f_1^2 \calv_t^2-432 c_1^2) H^4\\&+f_2^4
(11 f_1^2 \cala_t^2
+44 f_1^2 \cala_t \calv_t+44 f_1^2 \calv_t^2+45 c_1^2)\biggr)-\frac{c_3^2}{64 f_1^2 f_2^8 H^4 c_1} \biggl(
3 f_2^8 (5 f_1^2 \cala_t^2+20 f_1^2 \cala_t \calv_t\\&+20 f_1^2 \calv_t^2+3 c_1^2) H^8-48 f_2^6 H^6 f_1^2 c_1^2+(-30 f_1^2 f_2^8 \cala_t^2-120 f_1^2 f_2^8 \cala_t \calv_t-120 f_1^2 f_2^8 \calv_t^2\\&+16 f_2^4 f_1^4 c_1^2
-18 f_2^8 c_1^2+224 f_1^2 \calv_t^2+288 c_1^2) H^4+3 f_2^6 (5 f_1^2 f_2^2 \cala_t^2+20 f_1^2 f_2^2 \cala_t \calv_t\\&+20 f_1^2 f_2^2 \calv_t^2
-16 f_1^2 H^2 c_1^2+3 f_2^2 c_1^2)\biggr)\,;
\end{split}
\eqlabel{kwpap3}
\end{equation}
\begin{equation}
\begin{split}
&0=c_2''+\frac{13(c_2')^2}{8 c_2}+\biggl(
-\frac{c_3'}{c_3}+\frac{5f_2'}{2f_2}+\frac{5f_1'}{8f_1}+\frac{5c_1'}{8c_1}
\biggr) c_2'
+\biggl(-\frac{3c_2}{64f_2^4}-\frac{3c_2 H^4}{128f_2^4}-\frac{3c_2}{128H^4 f_2^4}\biggr)\\
&\times (K')^2
-\frac{c_2 (f_1^4 f_2^4-14)}{1288f_2^4 f_1^2 c_1^2} (\cala_t')^2
-\frac{c_2 (2 f_1^4 f_2^4-7)}{144f_2^4 f_1^2 c_1^2} (\calv_t')^2
-\frac{c_2 \cala_t' \calv_t' (f_1^4 f_2^4+7)}{72f_2^4 f_1^2 c_1^2}
-\frac{c_2c_1'}{8c_1} \biggl(\frac{4 f_2'}{f_2}\\&+\frac{f_1'}{f_1}\biggr)
+\frac{c_2(H')^2}{4H^2} -\frac{c_2 f_2'f_1'}{2f_1 f_2}
-\frac{3c_2(f_2')^2}{4f_2^2}
+\frac{9c_2 c_3^2K^4  ((\cala_t +2 \calv_t)^2 f_1^2-9 c_1^2)}{128c_1^2 f_2^8 f_1^2}
\\&+\frac{c_2 c_3^2 K^2}{128 H^4 f_2^8 f_1^2 c_1^2} \biggl(
5 f_1^2 f_2^4 (H^4+1)^2 \cala_t^2+4 \calv_t f_1^2 (H^4 (5 f_2^4 H^4+10 f_2^4-36)+5 f_2^4) \cala_t\\&
+4 f_1^2 (H^4 (5 f_2^4 H^4+10 f_2^4-72)+5 f_2^4) \calv_t^2-9 c_1^2 (H^4 (5 f_2^4 H^4+10 f_2^4-48)+5 f_2^4)\biggr)
\\&+\frac{c_2}{64 H^4 f_2^8 f_1^2 c_1^2} \biggl(
f_1^2 f_2^8 c_3^2 (H^4-1)^2 \cala_t^2+4 \calv_t f_1^2 f_2^8 c_3^2 (H^4-1)^2 \cala_t
+4 c_3^2 f_1^2 (H^4 (f_2^8 H^4-2 f_2^8\\&+72)+f_2^8) \calv_t^2
-c_3^2 c_1^2 (H^2 (9 f_2^8 H^6-48 f_1^2 f_2^6 H^4+16 f_1^4 f_2^4 H^2-18 f_2^8 H^2-48 f_1^2 f_2^6\\&+288 H^2)+9 f_2^8)\biggr)\,;
\end{split}
\eqlabel{kwpap4}
\end{equation}
\begin{equation}
\begin{split}
&0=f_1''-\frac{3f_1(f_2')^2}{4f_2^2} +\biggl(
\frac{7f_2'}{2f_2}+\frac{7c_1'}{8c_1}-\frac{c_3'}{c_3}+\frac{21c_2'}{8c_2}\biggr) f_1'
-\frac{3f_1 (H^4+1)^2(K')^2}{128f_2^4 H^4} 
\\&+\frac{(10 f_1^4 f_2^4-3)(\calv_t')^2}{48f_2^4 f_1 c_1^2} +\frac{\cala_t'\calv_t' (5 f_1^4 f_2^4+3)}{24f_2^4 f_1 c_1^2}
+\frac{(5 f_1^4 f_2^4-6)(\cala_t')^2}{96f_2^4 f_1 c_1^2}
-\frac{f_1f_2'}{2f_2} \biggl(
\frac{c_1'}{c_1}+\frac{3c_2'}{c_2}\biggr)
\\&+\frac{f_1(H')^2}{4H^2} -\frac{3f_1(c_2')^2}{8c_2^2} -\frac{3f_1 c_1'c_2'}{8c_2 c_1}
+\frac{9c_3^2K^4}{128f_2^8 f_1 c_1^2} \biggl((\cala_t+2 \calv_t)^2 f_1^2+7 c_1^2\biggr)
\frac{K^2 c_3^2}{128 f_2^8 f_1 c_1^2 H^4} \biggl(
\\&5 f_1^2 f_2^4 (H^4+1)^2 \cala_t^2+4 \calv_t f_1^2 (H^4 (5 f_2^4 H^4+10 f_2^4-36)+5 f_2^4) \cala_t+(4 f_1^2 H^4 (5 f_2^4 H^4\\&+10 f_2^4
-72)+20 f_1^2 f_2^4) \calv_t^2+3 H^4 c_1^2 (33 f_2^4 H^4+66 f_2^4-112)+99 f_2^4 c_1^2\biggr)\\&+\frac{c_3^2}{64 f_2^8 f_1 c_1^2 H^4}
\biggl(
f_1^2 f_2^8 (H^4-1)^2 \cala_t^2+4 \calv_t f_1^2 f_2^8 (H^4-1)^2 \cala_t+(4 f_1^2 H^4 (f_2^8 H^4-2 f_2^8\\&+72)
+4 f_1^2 f_2^8) \calv_t^2-H^2 c_1^2 (-135 f_2^8 H^6-48 f_1^2 f_2^6 H^4+272 f_1^4 f_2^4 H^2+270 f_2^8 H^2\\&-48 f_1^2 f_2^6-224 H^2)
+135 f_2^8 c_1^2\biggr)\,;
\end{split}
\eqlabel{kwpap5}
\end{equation}
\begin{equation}
\begin{split}
&0=f_2''+\frac{9(f_2')^2}{4 f_2} +\biggl(
\frac{f_1'}{2f_1}+\frac{c_1'}{2c_1}-\frac{c_3'}{c_3}+\frac{3c_2'}{2c_2}\biggr) f_2'
+\frac{5(H^4+1)^2(K')^2}{128f_2^3 H^4}
\\&-\frac{f_1^4 f_2^4+2}{288f_2^3 f_1^2 c_1^2} (\cala_t')^2
-\frac{2 f_1^4 f_2^4+1}{144f_2^3 f_1^2 c_1^2} (\calv_t')^2-\frac{\cala_t'\calv_t' (f_1^4 f_2^4-1)}{72f_2^3 f_1^2 c_1^2}
+\frac{f_2(H')^2}{4H^2}
-\frac{3f_2 c_2'}{8c_2} \biggl(
\frac{f_1'}{f_1}+\frac{c_1'}{c_1}
\biggr)
\\&-\frac{3f_2(c_2')^2}{8c_2^2}-\frac{f_2 f_1'c_1'}{8f_1 c_1}-\frac{7c_3^2 K^4}{128f_1^2 f_2^7 c_1^2} c_3^2 \biggl(
(\cala_t +2 \calv_t)^2 f_1^2-9 c_1^2\biggr)
-\frac{c_3^2 K^2}{128 f_1^2 f_2^7 H^4 c_1^2} \biggl(
\\&3 f_1^2 f_2^4 (H^4+1)^2 \cala_t^2+4 \calv_t f_1^2 (H^4 (3 f_2^4 H^4+6 f_2^4-28)+3 f_2^4) \cala_t+4 f_1^2 (H^4 (3 f_2^4 H^4\\&+6 f_2^4
-56)+3 f_2^4) \calv_t^2-3 c_1^2 (H^4 (9 f_2^4 H^4+18 f_2^4-112)+9 f_2^4)\biggr)+\frac{c_3^2}{64 f_1^2 f_2^7 H^4 c_1^2}
\biggl(\\&f_1^2 f_2^8 (H^4-1)^2 \cala_t^2+4 \calv_t f_1^2 f_2^8 (H^4-1)^2 \cala_t+(4 f_1^2 H^4 (f_2^8 H^4-2 f_2^8-56)+4 f_1^2 f_2^8) \calv_t^2\\
&-c_1^2 (H^2 (9 f_2^8 H^6+144 f_1^2 f_2^6 H^4-112 f_1^4 f_2^4 H^2-18 f_2^8 H^2+144 f_1^2 f_2^6-224 H^2)+9 f_2^8)\biggr)\,;
\end{split}
\eqlabel{kwpap6}
\end{equation}
\begin{equation}
\begin{split}
&0=H''-\frac{(H')^2}{H}+\biggl(
\frac{4 f_2'}{f_2}+\frac{f_1'}{f_1}+\frac{c_1'}{c_1}-\frac{c_3'}{c_3}+\frac{3c_2'}{c_2}\biggr) H'
-\frac{(H^8-1)(K')^2}{16f_2^4 H^3}+\frac{c_3^2 (H^8-1) K^2 }{16f_1^2 f_2^4 H^3 c_1^2}\\
&\times \biggl(
(\cala_t +2 \calv_t)^2 f_1^2-9 c_1^2\biggr)
+\frac{c_3^2 (H^4-1)}{8 f_2^2 f_1^2 H^3 c_1^2} \biggl(
(H^4+1) f_1^2 f_2^2 \cala_t^2+4 \calv_t f_1^2 f_2^2 (H^4+1) \cala_t\\&+4 (H^4+1)
f_1^2 f_2^2 \calv_t^2+3 H^2 c_1^2 (-3 f_2^2 H^2+8 f_1^2)-9 f_2^2 c_1^2\biggr)\,;
\end{split}
\eqlabel{kwpap7}
\end{equation}
\begin{equation}
\begin{split}
&0=K''+\biggl(
\frac{4 (H^4-1) H'}{(H^4+1) H}+\frac{f_1'}{f_1}+\frac{c_1'}{c_1}-\frac{c_3'}{c_3}+\frac{3c_2'}{c_2}\biggr) K'
+\frac{c_3^2 K}{f_2^4 f_1^2 c_1^2 (H^4+1)^2} \biggl(
f_1^2 (H^4 (f_2^4 H^4\\&+2 f_2^4+2 K^2)+f_2^4) \cala_t^2+4 \calv_t f_1^2
(H^4 (f_2^4 H^4+2 f_2^4+2 K^2-4)+f_2^4) \cala_t+(4 f_1^2 H^4 (f_2^4 H^4\\&
+2 f_2^4+2 K^2-8)+4 f_1^2 f_2^4) \calv_t^2
-3 c_1^2 (H^4 (3 f_2^4 H^4+6 f_2^4+6 K^2-16)+3 f_2^4)\biggr)\,;
\end{split}
\eqlabel{kwpap8}
\end{equation}
and a single first-order constrain,
\begin{equation}
\begin{split}
&0=(K')^2 (H^4+1)^2+32 f_2^4 H^2 (H')^2
-\frac{4H^4 (f_1^4 f_2^4+2) (\cala_t')^2}{9f_1^2 c_1^2}
-\frac{8H^4 (2 f_1^4 f_2^4+1) (\calv_t')^2}{9f_1^2 c_1^2}
\\&-\frac{16\cala_t' \calv_t' H^4 (f_1^4 f_2^4-1)}{9f_1^2 c_1^2}
-16 f_2^4 H^4 \biggl(\frac{3 (c_2')^2}{c_2^2}-\frac{(c_1')^2}{c_1^2}\biggr)
-16 f_2^4 H^4 \biggl(
\frac{4 f_2'}{f_2}+\frac{f_1'}{f_1}+\frac{c_1'}{c_1}\biggr) \biggl(
\frac{c_1'}{c_1}\\&+\frac{3c_2'}{c_2}\biggr)
-96 f_2^2 H^4 (f_2')^2-\frac{64 f_1'f_2' f_2^3 H^4}{f_1}
+\frac{H^4 K^4c_3^2}{c_1^2 f_1^2 f_2^4}  \biggl((\cala_t+2 \calv_t)^2 f_1^2-9 c_1^2\biggr)
+\frac{c_3^2 K^2}{f_2^4 f_1^2 c_1^2} \biggl(
\\&f_1^2 f_2^4 (H^4+1)^2 \cala_t^2+4 \calv_t f_1^2 (f_2^4 (H^4+1)^2-4 H^4) \cala_t+4 f_1^2 (H^4 (f_2^4 H^4+2 f_2^4-8)+f_2^4) \calv_t^2\\
&-3 c_1^2 (H^4 (3 f_2^4 H^4+6 f_2^4-16)+3 f_2^4)\biggr)+\frac{2 c_3^2}{f_2^4 f_1^2 c_1^2} \biggl(
f_1^2 f_2^8 (H^4-1)^2 \cala_t^2\\&+4 \calv_t f_1^2 f_2^8 (H^4-1)^2 \cala_t
+4 f_1^2 (H^4 (f_2^8 H^4-2 f_2^8+8)+f_2^8) \calv_t^2-c_1^2 (H^2 (9 f_2^8 H^6\\&-48 f_1^2 f_2^6 H^4+16 f_1^4 f_2^4 H^2-18 f_2^8 H^2-48 f_1^2 f_2^6+32 H^2)+9 f_2^8))\,.
\end{split}
\eqlabel{kwpapc}
\end{equation}
We verified that \eqref{kwpapc} is consistent with \eqref{kwpap1}-\eqref{kwpap8}.

\bibliographystyle{JHEP}
\bibliography{kwcharge2}

\providecommand{\href}[2]{#2}\begingroup\raggedright\begin{thebibliography}{10}

\bibitem{Klebanov:1998hh}
I.~R. Klebanov and E.~Witten, \emph{{Superconformal field theory on
  three-branes at a Calabi-Yau singularity}},
  \href{http://dx.doi.org/10.1016/S0550-3213(98)00654-3}{\emph{Nucl. Phys.}
  {\bf B536} (1998) 199--218},
  [\href{https://arxiv.org/abs/hep-th/9807080}{{\tt hep-th/9807080}}].

\bibitem{Maldacena:1997re}
J.~M. Maldacena, \emph{{The Large N limit of superconformal field theories and
  supergravity}}, \href{http://dx.doi.org/10.1023/A:1026654312961,
  10.4310/ATMP.1998.v2.n2.a1}{\emph{Int. J. Theor. Phys.} {\bf 38} (1999)
  1113--1133}, [\href{https://arxiv.org/abs/hep-th/9711200}{{\tt
  hep-th/9711200}}].

\bibitem{Douglas:1996sw}
M.~R. Douglas and G.~W. Moore, \emph{{D-branes, quivers, and ALE instantons}},
  \href{https://arxiv.org/abs/hep-th/9603167}{{\tt hep-th/9603167}}.

\bibitem{Kachru:1998ys}
S.~Kachru and E.~Silverstein, \emph{{4-D conformal theories and strings on
  orbifolds}}, \href{http://dx.doi.org/10.1103/PhysRevLett.80.4855}{\emph{Phys.
  Rev. Lett.} {\bf 80} (1998) 4855--4858},
  [\href{https://arxiv.org/abs/hep-th/9802183}{{\tt hep-th/9802183}}].

\bibitem{Buchel:2021yay}
A.~Buchel, \emph{{A bestiary of black holes on the conifold with fluxes}},
  \href{http://dx.doi.org/10.1007/JHEP06(2021)102}{\emph{JHEP} {\bf 06} (2021)
  102}, [\href{https://arxiv.org/abs/2103.15188}{{\tt 2103.15188}}].

\bibitem{Witten:1998zw}
E.~Witten, \emph{{Anti-de Sitter space, thermal phase transition, and
  confinement in gauge theories}},
  \href{http://dx.doi.org/10.4310/ATMP.1998.v2.n3.a3}{\emph{Adv. Theor. Math.
  Phys.} {\bf 2} (1998) 505--532},
  [\href{https://arxiv.org/abs/hep-th/9803131}{{\tt hep-th/9803131}}].

\bibitem{Candelas:1989js}
P.~Candelas and X.~C. de~la Ossa, \emph{{Comments on Conifolds}},
  \href{http://dx.doi.org/10.1016/0550-3213(90)90577-Z}{\emph{Nucl. Phys. B}
  {\bf 342} (1990) 246--268}.

\bibitem{Ceresole:1999zs}
A.~Ceresole, G.~Dall'Agata, R.~D'Auria and S.~Ferrara, \emph{{Spectrum of type
  IIB supergravity on AdS(5) x T**11: Predictions on N=1 SCFT's}},
  \href{http://dx.doi.org/10.1103/PhysRevD.61.066001}{\emph{Phys. Rev. D} {\bf
  61} (2000) 066001}, [\href{https://arxiv.org/abs/hep-th/9905226}{{\tt
  hep-th/9905226}}].

\bibitem{Turiaci:2023wrh}
G.~J. Turiaci, \emph{{New insights on near-extremal black holes}},
  \href{https://arxiv.org/abs/2307.10423}{{\tt 2307.10423}}.

\bibitem{Hartnoll:2008vx}
S.~A. Hartnoll, C.~P. Herzog and G.~T. Horowitz, \emph{{Building a Holographic
  Superconductor}},
  \href{http://dx.doi.org/10.1103/PhysRevLett.101.031601}{\emph{Phys. Rev.
  Lett.} {\bf 101} (2008) 031601}, [\href{https://arxiv.org/abs/0803.3295}{{\tt
  0803.3295}}].

\bibitem{Buchel:2009ge}
A.~Buchel and C.~Pagnutti, \emph{{Exotic Hairy Black Holes}},
  \href{http://dx.doi.org/10.1016/j.nuclphysb.2009.08.017}{\emph{Nucl. Phys. B}
  {\bf 824} (2010) 85--94}, [\href{https://arxiv.org/abs/0904.1716}{{\tt
  0904.1716}}].

\bibitem{Buchel:2017map}
A.~Buchel, \emph{{Singularity development and supersymmetry in holography}},
  \href{http://dx.doi.org/10.1007/JHEP08(2017)134}{\emph{JHEP} {\bf 08} (2017)
  134}, [\href{https://arxiv.org/abs/1705.08560}{{\tt 1705.08560}}].

\bibitem{Buchel:2018bzp}
A.~Buchel, \emph{{Klebanov-Strassler black hole}},
  \href{http://dx.doi.org/10.1007/JHEP01(2019)207}{\emph{JHEP} {\bf 01} (2019)
  207}, [\href{https://arxiv.org/abs/1809.08484}{{\tt 1809.08484}}].

\bibitem{Klebanov:2000hb}
I.~R. Klebanov and M.~J. Strassler, \emph{{Supergravity and a confining gauge
  theory: Duality cascades and chi SB resolution of naked singularities}},
  \href{http://dx.doi.org/10.1088/1126-6708/2000/08/052}{\emph{JHEP} {\bf 08}
  (2000) 052}, [\href{https://arxiv.org/abs/hep-th/0007191}{{\tt
  hep-th/0007191}}].

\bibitem{Buchel:2014hja}
A.~Buchel, \emph{{Effective Action of the Baryonic Branch in String Theory Flux
  Throats}}, \href{http://dx.doi.org/10.1007/JHEP09(2014)117}{\emph{JHEP} {\bf
  09} (2014) 117}, [\href{https://arxiv.org/abs/1405.1518}{{\tt 1405.1518}}].

\bibitem{Cassani:2010na}
D.~Cassani and A.~F. Faedo, \emph{{A Supersymmetric consistent truncation for
  conifold solutions}},
  \href{http://dx.doi.org/10.1016/j.nuclphysb.2010.10.010}{\emph{Nucl. Phys. B}
  {\bf 843} (2011) 455--484}, [\href{https://arxiv.org/abs/1008.0883}{{\tt
  1008.0883}}].

\bibitem{Minasian:1999tt}
R.~Minasian and D.~Tsimpis, \emph{{On the geometry of nontrivially embedded
  branes}}, \href{http://dx.doi.org/10.1016/S0550-3213(00)00035-3}{\emph{Nucl.
  Phys. B} {\bf 572} (2000) 499--513},
  [\href{https://arxiv.org/abs/hep-th/9911042}{{\tt hep-th/9911042}}].

\bibitem{Buchel:2006gb}
A.~Buchel and J.~T. Liu, \emph{{Gauged supergravity from type IIB string theory
  on Y**p,q manifolds}},
  \href{http://dx.doi.org/10.1016/j.nuclphysb.2007.03.001}{\emph{Nucl. Phys. B}
  {\bf 771} (2007) 93--112}, [\href{https://arxiv.org/abs/hep-th/0608002}{{\tt
  hep-th/0608002}}].

\bibitem{Zaffaroni:2000vh}
A.~Zaffaroni, \emph{{Introduction to the AdS-CFT correspondence}},
  \href{http://dx.doi.org/10.1088/0264-9381/17/17/306}{\emph{Class. Quant.
  Grav.} {\bf 17} (2000) 3571--3597}.

\bibitem{Buchel:2010wp}
A.~Buchel, \emph{{Chiral symmetry breaking in cascading gauge theory plasma}},
  \href{http://dx.doi.org/10.1016/j.nuclphysb.2011.01.031}{\emph{Nucl. Phys.}
  {\bf B847} (2011) 297--324}, [\href{https://arxiv.org/abs/1012.2404}{{\tt
  1012.2404}}].

\bibitem{Aharony:2005zr}
O.~Aharony, A.~Buchel and A.~Yarom, \emph{{Holographic renormalization of
  cascading gauge theories}},
  \href{http://dx.doi.org/10.1103/PhysRevD.72.066003}{\emph{Phys. Rev.} {\bf
  D72} (2005) 066003}, [\href{https://arxiv.org/abs/hep-th/0506002}{{\tt
  hep-th/0506002}}].

\bibitem{Buchel:2022zxl}
A.~Buchel, \emph{{The quest for a conifold conformal order}},
  \href{http://dx.doi.org/10.1007/JHEP08(2022)080}{\emph{JHEP} {\bf 08} (2022)
  080}, [\href{https://arxiv.org/abs/2205.00612}{{\tt 2205.00612}}].

\bibitem{Berenstein:2002ke}
D.~Berenstein, C.~P. Herzog and I.~R. Klebanov, \emph{{Baryon spectra and AdS
  /CFT correspondence}},
  \href{http://dx.doi.org/10.1088/1126-6708/2002/06/047}{\emph{JHEP} {\bf 06}
  (2002) 047}, [\href{https://arxiv.org/abs/hep-th/0202150}{{\tt
  hep-th/0202150}}].

\bibitem{Gubser:2009qm}
S.~S. Gubser, C.~P. Herzog, S.~S. Pufu and T.~Tesileanu, \emph{{Superconductors
  from Superstrings}},
  \href{http://dx.doi.org/10.1103/PhysRevLett.103.141601}{\emph{Phys. Rev.
  Lett.} {\bf 103} (2009) 141601}, [\href{https://arxiv.org/abs/0907.3510}{{\tt
  0907.3510}}].

\bibitem{Buchel:2024hdd}
A.~Buchel, \emph{{Plumbing the wormholes of string theory flux
  compactifications}},  \href{https://arxiv.org/abs/2404.00803}{{\tt
  2404.00803}}.

\bibitem{Aharony:2007vg}
O.~Aharony, A.~Buchel and P.~Kerner, \emph{{The Black hole in the throat:
  Thermodynamics of strongly coupled cascading gauge theories}},
  \href{http://dx.doi.org/10.1103/PhysRevD.76.086005}{\emph{Phys. Rev.} {\bf
  D76} (2007) 086005}, [\href{https://arxiv.org/abs/0706.1768}{{\tt
  0706.1768}}].

\bibitem{Balasubramanian:1999re}
V.~Balasubramanian and P.~Kraus, \emph{{A Stress tensor for Anti-de Sitter
  gravity}}, \href{http://dx.doi.org/10.1007/s002200050764}{\emph{Commun. Math.
  Phys.} {\bf 208} (1999) 413--428},
  [\href{https://arxiv.org/abs/hep-th/9902121}{{\tt hep-th/9902121}}].

\bibitem{Buchel:2019pjb}
A.~Buchel, \emph{{$\chi\rm{SB}$ of cascading gauge theory in de Sitter}},
  \href{http://dx.doi.org/10.1007/JHEP05(2020)035}{\emph{JHEP} {\bf 05} (2020)
  035}, [\href{https://arxiv.org/abs/1912.03566}{{\tt 1912.03566}}].

\end{thebibliography}\endgroup

\end{document}